\documentclass[pre,aps,showpacs,twocolumn]{revtex4-1}
\usepackage{graphicx}
\usepackage{amsmath}
\usepackage{amsfonts}
\usepackage{amssymb}
\usepackage{color}
\usepackage{bm}      
\usepackage{booktabs}
\usepackage{array}
\usepackage[english]{babel}
\usepackage{color}
\usepackage[bbgreekl]{mathbbol}  
\usepackage{slashed}  
\usepackage[hidelinks]{hyperref}
\hypersetup{pdfpagemode=FullScreen,  
colorlinks=true,
citecolor=blue} 
\usepackage{tikz}
\usepackage{verbatim}
\usepackage{notes2bib}
\bibnotesetup{
note-name = ,
use-sort-key = false
}

\newcommand{\bea}{\begin{eqnarray}}
\newcommand{\eea}{\end{eqnarray}}
\newcommand{\beq}{\begin{equation}}
\newcommand{\eeq}{\end{equation}}
\newcommand{\bit}{\begin{itemize}}
\newcommand{\eit}{\end{itemize}}
\newcommand{\ra}{\rightarrow}
\renewcommand{\r}{\right \rangle}
\renewcommand{\l}{\left \langle}
\newcommand{\D}{\mathrm{d}}

\newcommand{\s}{\sigma}
\newcommand{\e}{\epsilon}
\renewcommand{\u}{u}
\newcommand{\m}{m}
\newcommand{\bo}{\bm{\omega}}
\newcommand{\bk}{{\bm{k}}}
\newcommand{\bK}{\bm{K}}
\newcommand{\bcalK}{\bm{\mathcal{K}}}
\newcommand{\bpi}{\bm{\pi}}
\newcommand{\bkappa}{{\bm{\kappa}}}
\newcommand{\bgamma}{\bm{\gamma}}
 
\renewcommand{\P}{\mathcal{P}}

\usetikzlibrary{calc,trees,positioning,arrows,chains,shapes.geometric,%
    decorations.pathreplacing,decorations.pathmorphing,shapes,%
    matrix,shapes.symbols}

\tikzset{
>=stealth',
  punktchain/.style={
    rectangle, 
    rounded corners, 
    draw=black, very thick,
    text width=10em, 
    minimum height=3em, 
    text centered, 
    on chain},
  line/.style={draw, thick, <-},
  element/.style={
    tape,
    top color=white,
    bottom color=blue!50!black!60!,
    minimum width=8em,
    draw=blue!40!black!90, very thick,
    text width=10em, 
    minimum height=3.5em, 
    text centered, 
    on chain},
  every join/.style={->, thick,shorten >=1pt},
  decoration={brace},
  tuborg/.style={decorate},
  tubnode/.style={midway, right=2pt},
}

\begin{document}

\title{Nonequilibrium thermodynamic potentials for continuous-time Markov chains}
\author{Gatien Verley}
\affiliation{Laboratoire de Physique Th\'eorique (UMR8627), CNRS, Univ. Paris-Sud, Universit\'e Paris-Saclay, 91405 Orsay, France }
\date{\today}

\begin{abstract}
We connect the rare fluctuations of an Equilibrium (EQ) process and the typical fluctuations of a nonequilibrium (NE) stationary process. In the framework of large deviation theory, this observation allows us to introduce NE thermodynamic potentials. For continuous-time Markov chains, we identify the relevant pairs of conjugated variables and propose two NE ensembles: one with fixed dynamics and fluctuating time-averaged variables, and another with fixed time-averaged variables, but a fluctuating dynamics. Accordingly, we show that NE processes are equivalent to conditioned EQ processes ensuring that NE potentials are Legendre dual. We find a variational principle satisfied by the NE potentials that reach their maximum in the NE stationary state and whose first derivatives produce the NE equations of state, and second derivatives produce the NE Maxwell relations generalizing the Onsager reciprocity relations.
\end{abstract}

\maketitle

%%%%%%%%%%%%%%%%%%%%%%%%%%%%%%%%%%%
\section{Introduction}
%%%%%%%%%%%%%%%%%%%%%%%%%%%%%%%%%%%

Potentials define a specific concept in physics. They predict the evolution of a system from a variational principle. Such principles span many scientific fields from mechanics, electromagnetism and optics to control theory, thermodynamics and statistical physics. A variational principle elegantly summarizes the method used to solve a problem into the extremization of the appropriated cost function, for instance the action in mechanics \cite{Book_Landau1976}, the optical path length in optics \cite{Book_Born1999_vol}, or the thermodynamic potential in statistical physics \cite{Callen1985_vol}. The underlying idea is to explore all possibilities, including non physical ones, to find the physical solution from the extremum of the cost function. 

%%%%%		EQUILIBRIUM POTENTIALS 		%%%%%
In statistical physics, a thermodynamic potential is a state function of the thermodynamic variables. The latter specify a coarse-grained representation of the state of a system including a large number of degrees of freedom. Thermodynamic variables come in conjugated pairs: in each pair, one variable is free and one is constrained according to the environmental conditions. The EQ thermodynamic potentials proceed from the Legendre transformation of either energy or entropy. This transformation, at the core of the theory's dual structure, allows us to interchange the free and constrained variables. The thermodynamic state is reached at the extremum of the thermodynamic potential. There, the mean free variables are functions of the constrained ones. Beyond the mean description, the potential also predicts the statistics of the free thermodynamic variables, either by generating their cumulants, or from its connection with the asymptotic probability of the free variables.

%%%%%%			APPLICATIONS			%%%%%
Statistical physics provides a microscopic foundation to thermodynamics and a method to describe equilibrium systems. In the last decades, the large deviation theory \cite{Oono1989_vol99,Touchette2009_vol478} has modernized our understanding of statistical physics and accounted for its successes. More recently, it has received a growing interest thanks to its applications to NE systems, for instance in glasses \cite{Garrahan2009_vol42, Turci2011_vol94, Speck2012_vol109, Nemoto2014_vol2014}, biological systems \cite{Hertz1999_vol15, Lacoste2008_vol78, Ritort2008_vol137} or rare events sampling \cite{Giardina2011_vol145}.

%%%%%			LITERATURE				%%%%%
% Continous space processes
Clearly, one step toward understanding NE phenomena starts with the derivation of a NE thermodynamic potential verifying most of the aforementioned properties. With this in mind, many authors have shed light on the structure of statistical physics for NE Markov processes. Oono and Eyink considered that Large Deviation Functions (LDF) could represent NE potentials \cite{Oono1989_vol99, Eyink1996_vol54, Eyink1998_vol130}. On this basis, Oono and Paniconi proposed a phenomenological framework to study NE steady states \cite{Oono1998_vol}. For NE continuous processes, Bertini \textit{et al.} developed the macroscopic fluctuation theory describing the statistics of density and current fluctuations in Non-Equilibrium Stationary States (NESS) \cite{Bertini2002_vol107,Bertini2015_vol87}. Bodineau and Derrida used an additivity principle to predict those fluctuations in diffusive systems \cite{Bodineau2004_vol92,Derrida2007_vol}.

% Dicrete space processes
For discrete processes, Lecomte, Appert-Rolland and van Wijland introduced a dynamical partition function and the corresponding topological pressure identified as a LDF \cite{Lecomte2005_vol95, Lecomte2007_vol127}. Baule and Evans explored these ideas using a path entropy with the aim of finding rules constraining the dynamics of fluids under continuous shear \cite{Evans2010_vol51,Baule2008_vol101, Baule2010_vol2010}. Monthus proposed a similar approach, but involving the maximization of a trajectory-based relative entropy in the presence of constraints \cite{Monthus2011_vol2011}. Using large deviation theory, Maes and Neto{\v c}n{\'y} \cite{Maes2008_vol82} found a canonical structure and obtained the LDF of occupation and current probabilities from a variational approach based on the LDF of occupation and transition probabilities. A key step was the introduction of an EQ reference process to highlight that EQ fluctuations naturally appear when studying NE fluctuations. From another perspective, Nemoto and Sasa have shown that a Cumulant Generating Function (CGF) also proceeds from a variational principle, strengthening the dual structure of the theory \cite{Nemoto2011_vol}.

%%%%%%%%%%%%%%%%%%%%%%%%%%%%%%%%%%%%%%%
 \begin{center}
  \begin{table*}
    \caption{Relationship between the various stochastic processes and NE ensembles. The EQ reference process conditioned on the energy currents $j$, activities $f$ and occupations $p$ generates the trajectories of the systems in the NE micro-canonical ensemble. The process with mean energy currents, activities and occupations equal to the constrained values of the conditioned process is the driven process. The path probabilities of the driven process are asymptotically equivalent to the path probability of the NE process and of the canonical reference process. The CGF of $j,f$ and $p$ for the NE process is exactly the same as the CGF of the EQ reference process (up to a translation), i.e spontaneous rare fluctutions of the EQ process are associated to typical realizations of the NE process. The NE process generates the trajectories of the systems in the meta-canonical ensemble. This ensemble includes the systems that are put out-of-equilibrium by gradients of temperatures imposed by heat reservoirs. From the equivalence between the conditioned reference process with the driven reference process and the NE process, we conclude that the NE micro-canonical and meta-canonical ensembles are equivalent. \label{structure} 
  } 
   \begin{tikzpicture}
% THE DIAGRAM PART
  [node distance=2.0cm,
  start chain=going right, ]
  \node[punktchain, join, ]	(NeCanEns) {NE metacanonical \\ ensemble  $\Gamma(a,b,\m)$};
  \begin{scope}[start branch=venstre, every join/.style={->, thick, shorten <=1pt}, ]
    \node[punktchain, on chain=going below, node distance=1.0cm ] (EqProcess)		{EQ reference process $z$ with generator $\bk$};
    \begin{scope}[start branch=venstre, every join/.style={->, thick, shorten <=1pt}, ]
	  \node[punktchain, on chain=going below, node distance=1.0cm ]	(NeProcess) {NE process $\bar z$ with generator $\bar \bk$ };
	  \node[punktchain, on chain=going right, join ]	(CondNeProcess) {Conditioned NE process $\bar z | j,f,p$  };
	  \node[punktchain, on chain=going right, join ]	(DrivenNeProcess) {Driven NE process with generator $\bar \bK$};
    \end{scope}
    \node[punktchain, join, ] (CondEqProcess)	{Conditioned reference \\ process $z | j,f,p$} ;
    \node[punktchain, join, ] (DrivenEqProcess) 			{Driven reference \\ process with generator $\bK$} ;
  \end{scope} 
  \node[punktchain, ]	(NeMicroCanEns) {NE micro-canonical ensemble \\ $L(j,f,p)$}; 
  \node[punktchain, ]	(CanProcess) {Canonical reference \\ process with generator $\bcalK$  }; 
  \draw[|-,-|,<->, thick, ] (DrivenNeProcess.north) -- (DrivenEqProcess.south);
  \draw[|-,-|,<->, thick, ] (NeMicroCanEns.west) -- (NeCanEns.east);
  \draw[|-,-|,<->, thick, ] (DrivenEqProcess.east) --(13.5,-2.02)--(13.5,-5.3)--(0,-5.3)--  (NeProcess.south);
  \draw[|-,-|,<-, thick, ] (NeCanEns.west) -|(-2.5,-4.035)--  (NeProcess.west);
  \draw[|-,-|,<-, thick, ] (NeMicroCanEns.south) --  (CondEqProcess.north);
  \draw[|-,-|,<->, thick, ] (EqProcess.south) --  (NeProcess.north);
  \draw[|-,-|,->, thick, ] (EqProcess.north) --  (NeCanEns.south);
  \draw[|-,-|,<->, thick, ] (CanProcess.south) --  (DrivenEqProcess.north);
  %% THE PART TO ADD BRAKETS :
  %% No. 1
  \draw[tuborg] let
    \p1=(NeCanEns.west), \p2=(NeMicroCanEns.east) in
    ($(\x1,\y1+2.5em)$) -- ($(\x2,\y2+2.5em)$) node[above, midway]  {Ensembles of NE systems};
  %% THE PART TO ADD LABELS
\begin{scope}    
    \path (DrivenNeProcess) -- (DrivenEqProcess) node [midway, left] { Generators mapping };
    \path (NeMicroCanEns) -- (NeCanEns) node [midway, above] { Ensemble};
    \path (NeMicroCanEns) -- (NeCanEns) node [midway, below] { equivalence};
    \path (EqProcess.north) --  (NeCanEns.south) node [midway, right] {CGF built from the EQ process};
    \path (EqProcess) -- (CondEqProcess) node [midway, above] {  \begin{tabular}{c} \raisebox{-1ex}{Micro-} \\ \raisebox{-1ex}{canonical}
\end{tabular}};
    \path (EqProcess) -- (CondEqProcess) node [midway, below] { conditioning};
        \path (NeProcess) -- (CondNeProcess) node [midway, above] { \begin{tabular}{c} \raisebox{-1ex}{Micro-} \\ \raisebox{-1ex}{canonical}
\end{tabular}};
    \path (NeProcess) -- (CondNeProcess) node [midway, below] { conditioning};
    \path (CondNeProcess) -- (DrivenNeProcess) node [midway, above] { Optimization};
    \path (CondNeProcess) -- (DrivenNeProcess) node [midway, below] { problem};
    \path (CondEqProcess) -- (DrivenEqProcess) node [midway, above] { Optimization};
    \path (CondEqProcess) -- (DrivenEqProcess) node [midway, below] { problem};
    \path (EqProcess) -- (NeProcess) node [midway, right] { Tilted-operators mapping ($\bkappa \longleftrightarrow \bar \bkappa $)};
    \path (13.5,-5.3)--(0,-5.3) node [midway, below, sloped] { Similarity transformation $\bK= |\bpi| \cdot \bar \bk \cdot |\bpi|^{-1} \;\; \Rightarrow \;\;\; $Asymptotic path equivalence };
	\path (NeMicroCanEns.south) --  (CondEqProcess.north) node [midway, right] {Generates the path ensemble};
	\draw  (-2.5,-3.2)--(-2.5,-1) node [midway, above, sloped] {Generates the path ensemble};
    \path (CanProcess.south) --  (DrivenEqProcess.north) node [anchor=center, text width=3.5cm, auto, midway] {Asymptotic path equivalence};
\end{scope}
   \end{tikzpicture}
  \end{table*}
 \end{center}
%%%%%%%%%%%%%%%%%%%%%%%%%%%%%%%%%%%%%%%

% General framework
More recently, Chetrite and Touchette proposed a general framework for both continuous and discrete processes: they found that a conditioned Markov process is ensemble-equivalent to a condition-free process called the driven (or auxiliary) process, but also to an exponentially tilted process called the canonical process \cite{Chetrite2015_vol16,Chetrite2013_vol111}. This later process is defined by  exponentially weighting the probability of each trajectory with a weight depending on a functional $v$ of the stochastic process. This weighting procedure, is analogous to the definition of the canonical ensemble from the superposition of micro-canonical ensembles using a Boltzmann weight. On the other hand, the conditioned Markov process assumes that the variable $v$ is constrained to a given value. Finally, the driven process has a dynamics defined such that the mean value of $v$ is equal to the imposed value in the conditioned process. A systematic method of constructing this driven process from a variational approach was provided in Ref. \cite{Chetrite2015_vol2015}. A construction of the canonical process was also proposed by Giardin\'a, Kurchan and Peliti in Ref.~\cite{Giardina2006_vol96} for classical systems and by Garrahan and Lesanovsky in Ref.~\cite{Garrahan2010_vol104} for dissipative quantum systems. Jack and Sollich constructed a driven process for classical systems in Ref.~\cite{Jack2010_vol184}. The questions of the validity of the path ensemble equivalence has recently been studied in Ref.~\cite{Szavits-Nossan2015_vol2015}

%%%%%			MOTIVATIONS				%%%%%
Despite all these results, the structure of NE statistical physics is incomplete as regards to EQ statistical physics. For instance, the identification of the relevant coarse-grained degrees of freedom, i.e. the NE thermodynamic variables, is still missing. Accordingly, no general definition exists for stationary NE thermodynamic potentials. To progress in this direction, focusing on continuous-time Markov chains and stationary processes, we consider the following questions: can we describe the NE fluctuations of a system from the fluctuations of the \emph{same} system at EQ? If yes, can we define meaningful NE thermodynamic potentials using the variables involved in this correspondence? We positively answer these two questions by finding an exact mapping between the statistics of EQ and NE processes. This mapping involves, among others, the affinities of the NE process and some dynamical biases. The later parameter enables to dilate the energy barriers separating the various states of the system. The variables conjugated to the affinities and the dynamical biases are, respectively, the energy currents and the activities of the exchanges with the environment. The existence of a simple mapping when considering the appropriated couples of conjugated variables suggests that a complete canonical structure for NE statistical physics exists. With respect to previous works on conditioned Markov processes, our main contribution is to identify the constrains that does not modify the system dynamics, apart from changing the temperatures of the heat reservoirs. Accordingly, we define two ensembles of NE systems: the meta-canonical ensemble where the constrained variables are the affinities, and the NE micro-canonical ensemble where the constrained variables are the energy currents. We prove the equivalence of these ensembles and derive the NE thermodynamic potentials conjugated by Legendre transformation. We also obtain the NE equations of state connecting the conjugated variables.

%%%%%		TABLE OF CONTENTS			%%%%%
Our results and the structure of the theory are summarized in Table~\ref{structure}. Accordingly, the outline of the paper is as follows. We start by studying the fluctuations of an EQ reference process in Sec.~\ref{EQfluctuation} whose material corresponds to the middle row of Table~\ref{structure}. The definition of the EQ reference process and an introduction to large deviation theory are provided in Secs.~\ref{DefEQprocess} and \ref{LDtimeAVGvar}. After these introductory subsections, we look for an asymptotic approximation of the probability of the energy currents, activities and occupations of the systems states. Since we are dealing with an EQ system, no mean energy current exists. However, rare spontaneous fluctuations may produce non-zero energy currents and some arbitrary activities and occupations. We seek the probability of these events from an optimization problem: given that some energy currents $j$, activities $f$, and occupations $p$ are observed, defining the conditioned reference process, which process (called the driven reference process) reproduces these conditioned values $j$, $f$ and $p$ as typical values? We construct this driven process in Sec.~\ref{EQ-LDF} and obtain the LDF of $j$, $f$ and $p$. We use this result to derive the corresponding scaled CGF from a variational approach in Sec.~\ref{EQ-CGF}.

We switch to the study of the fluctuations of a NE process in Sec.~\ref{NEfluctuations}. This section corresponds to the third row of Table~\ref{structure}, which is obtained following exactly the same path as for the EQ reference process, except that we start with a NE process as defined in Sec.~\ref{DefNEprocess}: we look for the NE driven process that will typically reproduce the arbitrary energy currents $j$ and activities $f$ imposed in the NE conditioned process. Our first main result is to connect, in Sec.~\ref{MappingNEonEQ}, the EQ reference process and the NE process, and as a consequence, also to connect their associated driven processes (see the vertical arrows in Table~\ref{structure}). Our second main result is to prove, in Sec.~\ref{EnsembleEquivalence}, the asymptotic equivalence between the path probabilities of the driven reference process with the NE process. This equivalence is at the core of the aforementioned equivalence between the NE micro-canonical ensemble and the meta-canonical ensemble. In Sec.\ref{discussion}, we comment the structure of the theory starting with a short summary in Sec.~\ref{CanonicalStructure}. We discuss the symmetries of the NE potentials and the connection with close-to-equilibrium and far-from-equilibrium perturbation theory in Secs.~\ref{NEsymmetries} and \ref{NEresponse} respectively. We end by illustrating our work on a two-level model in Sec.\ref{Example}.

For the sake of simplicity, we focus on systems exchanging only energy with heat reservoirs. The generalization of our results to include matter, volume or other extensive variable exchanges with reservoirs is straightforward \cite{VandenBroeck2014_vol418}.

%%%%%%%%%%%%%%%%%%%%%%%%%%%%%%%%%%%
\section{Equilibrium fluctuations}
%%%%%%%%%%%%%%%%%%%%%%%%%%%%%%%%%%%
\label{EQfluctuation}

\subsection{Definition of the EQ reference process}
\label{DefEQprocess}

\begin{figure}
\includegraphics[width=6.5cm]{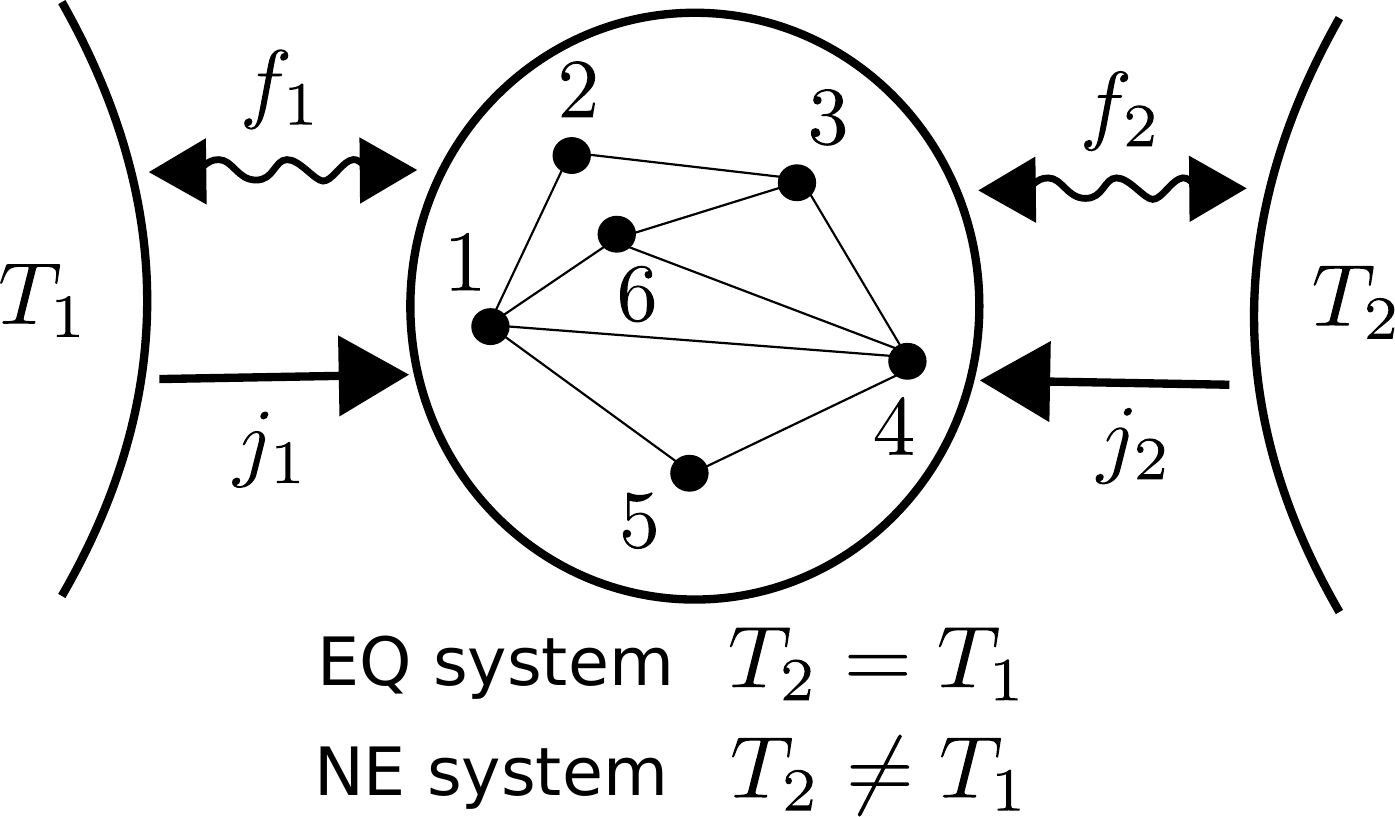}
\caption{System with $M=6$ states connected to $\chi = 2$ heat reservoirs at the same temperature $T_1 = T_2 $ for the EQ reference process, or at different temperature $T_1 \neq T_2$ for the NE process. \label{fig10}}
\end{figure}

%%%%% Global parameters
We consider an \emph{EQ reference process} corresponding to a physical system modeled by a continuous-time Markov chain with a finite number $M$ of discrete states. This system exchanges energy with $\chi $ heat reservoirs labeled by $\nu =1 \dots \chi$ at inverse temperatures $ \beta_1 = 1/ (k_B  T_1)$, with $ k_B =1 $ the Boltzmann constant, see Fig.~\ref{fig10}. The reference process is at EQ, i.e. all the heat reservoirs share the same inverse temperature $\beta_1 $. We use several heat reservoirs to allow different mechanisms of energy exchange. As a result, some rare events with net energy flow from one heat reservoir to another may occur. The system states are generically denoted $x$, $y$ and $z$. The state at time $\tau$ is $z(\tau)$. A system state trajectory during time interval $[0,t]$ is denoted $[z]$. This trajectory includes the state $z(\tau)$ at all time $\tau \in [0,t]$ and the label $\nu(\tau)$ of the reservoir providing the energy at each change of state in the trajectory. 

\begin{figure}
\includegraphics[width=7cm]{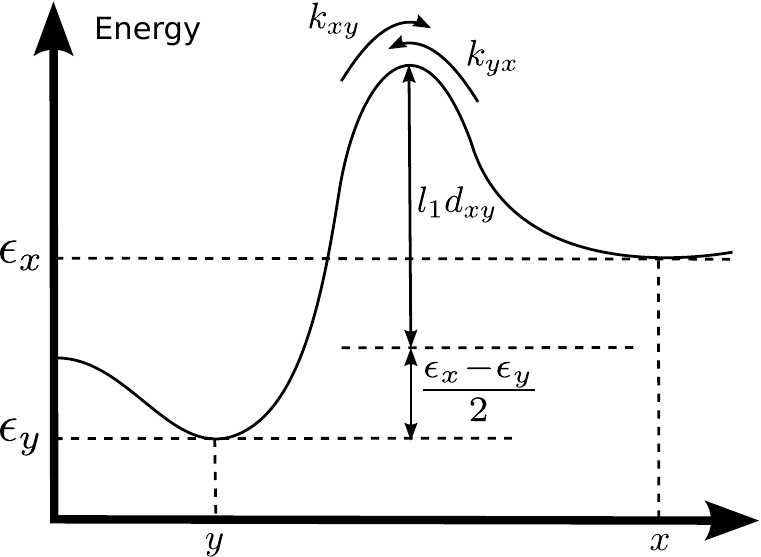}[h]
\caption{Energy lanscape for the $x \leftrightarrow y$ transition. The discrete states $x$ and $y$ represent the locations of the minima in the energy landscape. Changing the dilatation factor $l_1$ modify the height of all energy barriers for the EQ reference process.  \label{fig11}}
\end{figure}

The energy of state $x$ is $\e_x$. The probability per unit time of switching from state $y$ to state $x$ exchanging the energy $\e_x-\e_y$ with reservoir $\nu$ is given by the Arrhenius transition rates
\begin{equation}
 k_{xy}^{\nu} \equiv \gamma_{xy}^{\nu} e^{-\beta_1 (\e_x-\e_y)/2 - \beta_1 l_1 d_{xy}}. \label{defEQrates}
\end{equation}
We have introduced the symmetric matrices $ \bgamma^{\nu} $, whose $(x,y)$ element yields the coupling with reservoir $ \nu $ for a transition from $ y $ to $ x $. The $(x,y)$ element of the symmetric matrix $\bm{d}$ represents the height of the energy barrier that must be crossed when the system switches between states $y$ and $x$, see Fig.~\ref{fig11}. The dimensionless parameter $l_1$ is a \emph{dilatation factor} that enables to modify the height of the energy barriers ($l_1 =1$ implies no dilatation).
The transition rates defined in Eq.~(\ref{defEQrates}) verify for all $\nu$ the local detailed balance relation
\begin{equation}
\ln \frac{k^{\nu}_{xy}}{k^{\nu}_{yx}} = - \beta_1(\e_x-\e_y),
\end{equation}
which ensures that the system will reach EQ \cite{VanKampen2007_vol}.
The reference probability per unit time of escaping from state $y$, given that energy is exchanged with reservoir $\nu$, is denoted
\begin{equation}
 \lambda^{\nu}_y \equiv \sum_{x\neq y}  k^{\nu}_{xy} = - k^{\nu}_{yy},
\end{equation}
such that each column of  the matrix $ \bk^{(\nu )}$ sums to zero as required for continuous time Markov chains. The reference transition rate matrix $ \bk \equiv \sum_\nu  \bk^{(\nu )}$ returns the transition probabilities per unit time disregarding the reservoir involved in the energy exchanges. Similarly, $ \lambda \equiv \sum_\nu  \lambda^{(\nu )} $ is the total escape-rate vector. As a convention, we drop the subscripts of vector or matrix elements to refer to the whole vector or matrix and use bold face letters for matrices. We denote the ensemble average over all trajectories $[z]$ generated with dynamics corresponding to $ \bk$ with the brackets $\l \dots \r_{\bk} $.

\subsection{Large deviations of empirical time-averaged variables}

\label{LDtimeAVGvar}

%%%%% Basic fact on Large deviations
Throughout the paper, we assume that the long-time statistics of time-extensive variables obey a large deviation principle. For instance, let $z(t)$ be the position at time $t$ of a random walker on a one-dimensional circular lattice and $X[z]$ the number of steps the walker takes during the trajectory $[z]$. We remark that the variable $X$ is a functional of the trajectory $[z]$ that is a realization of a stochastic process. $X$ is not a random variable in itself. When $X$ is not evaluated on a trajectory, it refers either to the physical variable ``number of step'' or to a numerical value of this variable. The variable $X[z]$ is \emph{time-extensive} since $X[z]+X[z']=X[z,z']$, where $[z,z']$ denote the trajectory made with $[z]$ followed by $[z']$. Then, the number of transitions typically increases with time. Accordingly, $v[z]=X[z]/t$ is the number of steps per unit time and is regarded as a \emph{time-averaged} variable. At long time, it converges to the step frequency of the walker. The probability of $ v[z] = v $, i.e. that the time-averaged number of steps $v[z]$ takes the value $v$ at long time $t$, is $P_t (v) \simeq e^{-t I(v)}$.
The function $I$ is called a large deviation function (LDF). It is non-negative and vanishes at $v = \l v[z] \r_{\bk}$, denoting that
the ensemble average value is the most likely time-averaged $v$. Small (respectively large) deviations correspond to the time-averaged number of steps that are close-to (respectively far-from) the ensemble average value. These events become exponentially unlikely with increasing time for ergodic systems. The convexity of the LDF ensures that a large deviation is less likely than a small fluctuation.

%%%%% Definitions of relevant variables 
Following, we introduce the empirical time-averaged variables used to derive our central results. We name \emph{empirical variables} those that are defined from experimental observations of the system and that usually depend on the observed trajectory $[z]$.
First, we define the empirical occupation in $x$ by
\begin{equation}
p_x[z] \equiv  \frac{1}{t} \int_0^t \D \tau \delta_{x,z(\tau)},
\end{equation}
where $\delta$ is the Kronecker symbol. Given the probability of each state being gathered into the column vector $ p=(p_1, \cdots, p_M)^\dag $, the Shannon entropy $s =s(p)$ is
\begin{equation}
s(p)\equiv - \sum_x p_x \ln p_x = - ( p^\dag \cdot \ln p) , \label{def:AvgS}
\end{equation}
and the energy $e=e(p)$ is
\begin{equation}
e(p) \equiv \sum_x  \e_x p_x \label{def:AvgE} = \e^\dag \cdot p,
\end{equation}
with the central dot denoting the matrix product and $\dag$ the transposition. 
The time-averaged energy along trajectory $ [z] $ can be written
$ e[z] = e( p[z]) $, and similarly for the entropy. % if we denote $\circ$ the composition of two functions.
Second, we define the empirical transition probability from $y$ to $x$ induced by reservoir $\nu$
\begin{equation}
\omega^{\nu}_{xy}[z] \equiv \frac{1}{t} \sum_{\tau \in [0,t]} \delta_{x,z(\tau+d\tau)}\delta_{y,z(\tau)} \delta_{\nu, \nu(\tau)},
\label{def:TimeAvgOmega}
\end{equation}
where the sum is over all time $\tau$ at which the system changes from state $z(\tau)$ to state $ z(\tau+d\tau) $, exchanging energy with reservoir $ \nu(\tau) $. Given a transition probability $ \omega^{\nu}_{xy} $ for each possible transitions, the current of energy received from reservoir $\nu$ by the system is
\begin{equation}
j_\nu(\bo) \equiv \frac{1}{2}\sum_{x,y} \left( \omega^{\nu}_{xy} -\omega^{\nu}_{yx} \right)(\e_x-\e_y).
\label{def:TimeAvgJ}
\end{equation}
Its empirical value during trajectory $ [z] $ is written $ j_\nu[z] = j_\nu (\bo[z]) $. These time-averaged currents describe the anti-symmetric part of fluctuations since they change sign upon time-reversal of the trajectories. On the contrary, the weighted frequency of interaction with reservoir $ \nu $, named \emph{activity} for short and written
\begin{equation}
f_\nu(\bo) \equiv \frac{1}{2}\sum_{x,y\neq x} \left( \omega^{\nu}_{xy} +\omega^{\nu}_{yx} \right )d_{xy},
\label{def:TimeAvgF}
\end{equation}
describes the symmetric part of fluctuations. Indeed, $f_\nu[z] = f_\nu ( \bo[z] )$ does not change sign upon time-reversal of the trajectory $[z]$.
When the activity is low (high), the system either changes of state less (more) frequently or mostly switches between states with low (high) $d_{xy}$.
The term ``activity'' was proposed to qualify the symmetric part of the fluctuations in \cite{Lecomte2007_vol127, Baiesi2009_vol103, Baiesi2009_vol137, Baerts2013_vol88, Polettini2015_vol48}, see also references therein. Let us finally remark that, in the definitions of the energy currents and activities, the one-half factor is just a symmetry factor since we can sum over transitions disregarding their directions ($\sum_{x,y}$) or for only one direction ($\sum_{x>y}$). Half of the first sum is equivalent to the second sum. 

%%%%%%%%%%%%%%%%%%%%%%%%%%%%%%%%%%%
\subsection{LDF of energy currents, activities and occupation from a variational approach}
%%%%%%%%%%%%%%%%%%%%%%%%%%%%%%%%%%%
\label{EQ-LDF}

At long time $ t $, the probability of observing an empirical transition probability $\bo[z]=\bo$ and an empirical occupation $p[z]=p$ is
\begin{equation}
P_t(\bo,p ) \underset{t \ra \infty}{\simeq} e^{-t I(\bo,p)}. \label{AsymPJumpOccup}
\end{equation}
From the work of Maes and Neto$\check{c}$n\'y \cite{Maes2008_vol82}, Wynants \cite{Wynants2010_vola} or Bertini \textit{et al.} \cite{Bertini2015_vol51}, the LDF $I(\bo,p)$ of the empirical transition probabilities and occupations for the continuous-time Markov chain with generator $\bk$ is
\begin{equation}
I(\bo,p) = \sum_{x,y \neq x, \nu} \left[   k^{\nu}_{xy}p_y  - \omega^{\nu}_{xy} + \omega^{\nu}_{xy}  \ln \frac{\omega^{\nu}_{xy} }{ k^{\nu}_{xy} p_y}  \right ],
\label{LDF:OccupJump}
\end{equation}
where the sum is over $\nu$ from $1$ to $\chi$ and all couples $(x,y)$ such that $x \neq y$. 
The derivation of Eqs.~(\ref{AsymPJumpOccup}-\ref{LDF:OccupJump}) is reproduced in Appendix \ref{LDFoccupationandjump}. 

In Ref.~\cite{Maes2008_vol82}, the LDF of the occupation and probability current was obtained from a constrained optimization problem constructed with $I(\bo,p)$. This procedure, called ``contraction'' \cite{Oono1989_vol99}, is equivalent at the level of probabilities to marginalize $P_t(\bo,p)$ to obtain the probability of currents and occupations. We now proceed to the contraction of $I(\bo,p)$ to obtain  the LDF of energy currents, activities and occupations denoted $L(j,f,p)$. The long-time asymptotic approximation of the probability $P_t(j,f,p)$ that $j[z]=j$, $f[z]=f$ and $p[z]=p$ at time $t$ will then read
\begin{equation}
P_t(j,f,p) \underset{t \ra \infty}{\simeq} e^{ -t L(j,f,p)}. \label{ProbaAsymptotic}
\end{equation}
We prove in Appendix \ref{AppendixCountStat} a sharper approximation of this probability that involves a pre-exponential factor dictating the thermodynamic behavior: it leads to the statistics of the usual EQ thermodynamic variables that only depend on the system state, such as energy for instance. It is the first correction to the exponential decay of non-typical time-extensive variables after a long time. This prefactor was first obtained in Ref. \cite{Polettini2015_vol48}, but we provide in Appendix \ref{AppendixCountStat} a logically independent derivation (though restricted to the large time limit) that involves some results of Sec.\ref{NEfluctuations}.

At long time, the energy current $j $, activity $f$ and occupation $p$ mainly appear thanks to the most likely event producing them. The probability of this event is associated with smaller values of $I(\bo,p)$ with $\bo$ constrained by the value of the energy currents and activity. For this reason, and in virtue of the contraction principle, we minimize $I(\bo,p)$ under the energy currents constraint
\begin{equation}
j_\nu=j_\nu(\bo), \label{CurrentConstraints}
\end{equation}
for $\nu > 1$, because current conservation imposes $j_1 = -\sum_{\nu\neq 1} j_\nu $.  We also impose the activity constraint 
\begin{equation}
f_\nu=f_\nu(\bo). \label{TrafficConstraints}
\end{equation}
In addition, the probability currents should be compatible with the conservation of the norm of the occupation vector, i.e. for all $y$
\begin{equation}
\sum_{x ,\nu} \left(\omega^{\nu}_{xy}- \omega^{\nu}_{yx}\right) = 0 \label{ConsNormProba}.
\end{equation}
To perform our optimization problem, we use the following cost function
\begin{multline}
\mathcal{F}(\bo,p) = I(\bo,p) +  \sum_{\nu } a_\nu [j_\nu - j_\nu(\bo)] \\
 + \sum_{\nu } b_\nu  [f_\nu - f_\nu(\bo)] + \sum_{x,y,\nu}   \u_y \left( \omega^{\nu}_{xy}- \omega^{\nu}_{yx}  \right)  \label{OptimizationFunctional}
\end{multline}
where $a_\nu$, $ b_\nu $ and $ \u_y $ are Lagrange multipliers that will be chosen to satisfy the constraints of Eqs.~(\ref{CurrentConstraints}-\ref{ConsNormProba}). 
We choose $a_1= 0$ so as not to constrain the current $j_1$ that is already set by the current conservation law. We now minimize the function $\mathcal{F}$ with respect to $\bo$, calculating $\partial \mathcal{F} / \partial \omega^{\nu}_{xy} = 0$ to get
\begin{equation}
0 =  \ln \frac{\omega^{\nu}_{xy} }{ k^{\nu}_{xy} p_y} - a_\nu(\e_x-\e_y) -  b_\nu d_{xy} + (\u_y-\u_x), \label{eq:OptimJump}
\end{equation}
where we have used Eq.~(\ref{LDF:OccupJump}) and Eqs.~(\ref{CurrentConstraints}-\ref{TrafficConstraints}). Therefore, the optimal transition probability in terms of the Lagrange multipliers satisfies
\begin{equation}
\omega^{\nu}_{xy} =  K^{\nu}_{xy} p_y, \label{def:BiasedRates}
\end{equation}
where we have introduced $ \bK^{\nu} = \bK^{\nu}(a,b,\u) $, the transition probability for mechanism $ \nu $ divided by the empirical occupation of the state before transition. Its off-diagonal elements are
\begin{equation}
 K^{\nu}_{xy} \equiv  k^{\nu}_{xy} e^{a_\nu (\e_x-\e_y)+b_\nu d_{xy} + \u_x-\u_y} , \label{BiasRate} 
\end{equation}
or more explicitly using Eq.~\ref{defEQrates}
\begin{equation} 
 K^{\nu}_{xy} = \gamma_{xy}^{\nu} e^{-(\beta_1/2-a_\nu) (\e_x-\e_y) - ( \beta_1 l_1 - b_\nu ) d_{xy}+ \u_x-\u_y}, \label{ExplicitBiasRate}
\end{equation}
and the diagonal elements are
\begin{equation}
 K^{\nu}_{yy} = - \sum_{x\neq y}  K^{\nu}_{xy} \equiv -  \Lambda^{\nu}_y  , \label{DiagBiasRate}
\end{equation}
such that any column of any matrix $  \bK^{\nu} $ sums to zero. We remark that the matrices $ \bK^{\nu}$ satisfy a modified detailed-balance relation
\begin{equation}
\ln \frac{ K^{\nu}_{xy}}{ K^{\nu}_{yx}} = (2a_\nu- \beta_1)(\e_x-\e_y) + 2(\u_x-\u_y). \label{NEdetailbalance}
\end{equation}
In this local detailed balance, the Lagrange multiplier $a_\nu$ biases the inverse temperatures $\beta_1$ to make typical the energy exchanges corresponding to the energy currents constraint. The reservoir $\nu$ behaves as if it had the temperature $\beta_\nu \equiv \beta_1-2a_\nu$ in order to satisfy the current constraint. Thus, the variable 
\begin{equation}
2 a_\nu = \beta_1-\beta_\nu
\end{equation} is an \emph{affinity} \cite{Book_Prigogine1955,Book_Donder1927,Tome2015_vol91}, also called \emph{thermodynamic force} \cite{Andrieux2007_vol127,Andrieux2007_vol}. Notice that $a_1=0$ as required. The similarity between Eqs.~(\ref{defEQrates}) and (\ref{ExplicitBiasRate}) indicates that we can also introduce new  dilatation factors $l_\nu$ such that the \emph{dynamical bias} 
\begin{equation} 
b_\nu \equiv \beta_1 l_1 -\beta_\nu l_\nu
\end{equation} 
gives the modification of the dynamics in order to satisfy the activity constraint. Finally, we call the variable $\u$ the \emph{drift} because it acts like a force biasing each transition.

The explicit solution $ \bo $ of our variational problem $ \D \mathcal{F} = 0 $ is now almost reached. The next step is to use the constraints of Eqs.~(\ref{CurrentConstraints}-\ref{ConsNormProba}) to obtain the Lagrange multipliers. More explicitly the constraint equations are
\bea
j_{\nu}&=& \frac{1}{2}\sum_{x,y} \left(  K^{\nu}_{xy}p_y - K^{\nu}_{yx}p_x \right)(\e_x-\e_y), \label{EqOfStatej} \\
f_{\nu}&=& \frac{1}{2}\sum_{x,y} \left(  K^{\nu}_{xy}p_y + K^{\nu}_{yx}p_x \right)d_{xy}, \label{EqOfStatef} \\
0 &=&  \bK \cdot p, \label{StationaryEq}
\eea
where $\bK = \sum_\nu \bK^{\nu}$ is the generator of the \emph{driven reference process} \cite{Chetrite2015_vol16,Chetrite2015_vol2015}.
For the third equation, the conservation law of the probability current of Eq.~(\ref{ConsNormProba}) is reformulated as a requirement that the empirical occupation $p$ is the stationary probability of the continuous-time Markov chain with rate matrix $\bK= \bK(a,b,\u)$. Inverting these three equations gives the vectors $a$, $b$ and $u$ as a function of $(j,f,p)$. 

The final step to obtain the asymptotic probability of energy currents, activities and occupations is to write the LDF of Eq.~(\ref{LDF:OccupJump}) at the optimal transition probability of Eq.~(\ref{def:BiasedRates}). This leads to 
\begin{equation}
L(j,f,p) =   a^\dag   \cdot j  + b^\dag \cdot f + \left(  \lambda-  \Lambda \right)^\dag \cdot p ,    \label{LDFExtensive}
\end{equation}
where we have used the anti-symmetry of $\e_x-\e_y$ or symmetry of $d_{xy}$ in the exchange of $x$ and $y$ to make explicit the dependence in $ j $ and $ f $. We also used Eq.~(\ref{ConsNormProba}) to get rid of the term involving $\u_x-\u_y$.

\subsection{Scaled CGF of energy currents, activities and occupations from a variational approach}

\label{EQ-CGF}

In the previous section, we have obtained the LDF $L$ from the solution of an optimization problem. From now on, and for the remainder of the paper, we assume the convexity of the LDF. Our aim here is to derive the scaled CGF conjugated to $L$ from a variational approach, using the fact that LDF and scaled CGF are conjugated by Legendre transformation \cite{Oono1989_vol99,Touchette2009_vol478,Chetrite2015_vol2015}. On the way, we obtain useful properties associated to the canonical structure.

The scaled CGF of the energy current, activity and occupation is defined by 
\begin{equation}
\Gamma(a',b',\m') \equiv \lim_{t \rightarrow \infty} \frac{1}{t} \ln \l e^{t(a'^\dag\cdot j[z] + b'^\dag\cdot f[z] + \m'^\dag \cdot p[z])} \r_{\bk}, \label{DefCGF}
\end{equation}
and is the Legendre transformation of $L$
\begin{equation}
\Gamma(a',b',\m') = \max_{p,j,f} \left[ a'^\dag \cdot j + b'^\dag \cdot f + \m'^\dag \cdot p - L(p,j,f) \right ].
\end{equation}
The maximum on $j$ and $f$ is reached for $a'=a$ and $b'=b$, and the scaled CGF becomes
\begin{equation}
\Gamma(a,b,\m') = \max_{ p \,|\, \bK \cdot p =0} \left[ (\m'+  \Lambda -  \lambda)^\dag \cdot p \right],
\end{equation}
where the maximum is taken over all occupations with the Lagrange multiplier $\u$ in the generators of the driven process $ \bK$ tuned such that $ \bK \cdot p =0$. An alternative expression of the scaled CGF of energy current, activity and occupation is
\begin{equation}
\Gamma(a,b,\m') = \max_{ u } \left[ (\m'+ \Lambda- \lambda)^\dag \cdot p \right] \label{VariationalCGF}
\end{equation}
with $p$ the stationary probability associated to $ \bK$. From the optimal drift $\u=\u(a,b,\m')$ realizing the maximum in Eq.~(\ref{VariationalCGF}), we introduce the \emph{escape weight} $\m\equiv\m(a,b,\u)$ giving the value of $\m'$ for given $(a,b,\u)$. In Eq.(G14) of the appendix of reference \cite{Nemoto2011_vol}, Nemoto and Sasa gave the scaled CGF of energy current from a variational expression analogous to our Eq.~(\ref{VariationalCGF}). We recover their result taking $b =0$ and $\m'=0$. We further comment Eq.~(\ref{VariationalCGF}) noticing that the maximum is reached for $\u$ satisfying
\begin{equation}
 \Gamma(a,b,\m)= \m_y +  \Lambda_y -  \lambda_y \label{exitrule},
\end{equation}
for all $y$. This equation allows us to derive the following escape-rate rule
\begin{equation}
\m_y +  \Lambda_y -  \lambda_y = \m_x +  \Lambda_x -  \lambda_x,
\end{equation}
that can be related to the exit rate constraint of Refs.~\cite{Chetrite2015_vol16, Baule2008_vol101, Baule2010_vol2010} taking $\m = 0$. To prove Eq.~(\ref{exitrule}), we introduce the \emph{tilted operator} $\bkappa =\bkappa(a,b,\m) $ for the EQ reference process
\bea
 \kappa_{yy} &\equiv& -\sum_{x\neq y, \nu}  k^{\nu}_{xy} + \m_y, \label{DressedOperatorDiag}\\
 \kappa_{xy} &\equiv&  \sum_{\nu}  k^{\nu}_{xy} e^{a_\nu(\e_x-\e_y)+b_\nu d_{xy}}  .  \label{DressedOperator}
\eea
The generator of the driven reference process $ \bK$ is connected to this tilted operator by
\begin{equation}
 K_{xy} = e^{\u_x}  \kappa_{xy} e^{-\u_y}
- \left(m_y  +  \Lambda_y- \lambda_y \right) \delta_{xy}. \label{kappaToK}
\end{equation}
Using this equation and $  \bK \cdot p = 0$, we find
\begin{equation}
\sum_y e^{\u_x} \kappa_{xy} e^{-\u_y} p_y = \left(\m_x +  \Lambda_x- \lambda_x \right) p_x,
\end{equation}
after summing over $x$ and maximizing over $\u$, it follows from Eq.~(\ref{VariationalCGF}) that the drift giving the maximum satisfies
\begin{equation}
\sum_{x,y} e^{\u_x}  \kappa_{xy} e^{-\u_y} p_y = \Gamma.
\end{equation}
By definition \cite{Lebowitz1999_vol95}, the scaled CGF $ \Gamma(a,b,\m)$ is the highest eigenvalue of $ \bkappa$. Then $\pi_x \equiv \pm e^{\u_x} / Z(\u) $ is the normalized left eigenvector of $\bkappa$ with $Z(\u)$ a normalization constant such that $ \sum_x \pi_x = 1$. 
The vector $r \equiv \bpi^{-1} \cdot p $ is a right eigenvector with $\bpi_{xy} \equiv \pi_x \delta_{xy}$. Its norm is set by $\sum_x \pi_x r_x = \sum_x p_x = 1$. Notice that we cannot determine from the values of $\u$ the sign of each component of the vectors $\pi$ and $r$, but their $x$ components share the same sign. Now, summing Eq.~(\ref{kappaToK}) over $x$ leads to Eq.~(\ref{exitrule}) since $\sum_x K_{xy} = 0$ and $\sum_x e^{\u_x} \kappa_{xy} e^{-\u_y} = \Gamma$. 

Then, the optimal drift $\u = \u(a,b,\m)$, leading to the maximum in Eq.~(\ref{VariationalCGF}), is simply obtained from the left eigenvector of the tilted operator by $ \ln |\pi_x| = u_x - \ln |Z(\u)|$ up to a constant that plays no role, since only differences of drifts matter. The drift makes the escape-rate rule holds true and, using Eqs.~(\ref{LDFExtensive}) and (\ref{exitrule}), leads to the Legendre structure that one expects for LDFs and scaled CGFs. Finally, from Eqs.~(\ref{exitrule}) and (\ref{kappaToK}), we recover the results of Refs.~\cite{Jack2010_vol184, Chetrite2015_vol16, Chetrite2013_vol111, Chetrite2015_vol2015} in which the generator $\bK$ of the driven process corresponds to the Doob's transformation of the tilted operator $\bkappa$
\begin{equation}
 K_{xy} = |\pi_x|  \kappa_{xy} |\pi_y|^{-1}
- \Gamma \delta_{xy}.
\end{equation}
Notice that in Refs.~\cite{Chetrite2015_vol16, Chetrite2013_vol111, Chetrite2015_vol2015}, the right eigenvector of the tilted operator is used in the Doob's transformation instead of the left one, since the tilted operator in these references is the adjoint of $\bkappa$.

%%%%%%%%%%%%%%%%%%%%%%%%%%%%%%%%%%%
\section{Non-equilibrium fluctuations}
%%%%%%%%%%%%%%%%%%%%%%%%%%%%%%%%%%%
\label{NEfluctuations}

\subsection{Definition of the NE process}

\label{DefNEprocess}

The \emph{NE process} is defined by the rate matrices $ \bar \bk^{\nu} = \bar \bk^{\nu}(a_\nu,b_\nu)$ associated to energy exchanges with each reservoir $\nu$ at different temperatures $\beta_\nu =\beta_1 -2a_\nu $ and with different dilatation factors $l_\nu$ related to dynamical bias by $b_\nu = \beta_1 l_1 - \beta_\nu l_\nu$. The elements of the rate matrices are
\begin{equation}
\bar k_{xy}^{\nu} \equiv \gamma_{xy}^{\nu} e^{-(\beta_1/2-a_\nu) (\e_x-\e_y) - (\beta_1 l_1 - b_\nu) d_{xy}} \label{defNErates}.
\end{equation}
Accordingly, the escape rate $\bar  \lambda^{\nu}_y= \bar \lambda^{\nu}_y(a_\nu,b_\nu)$ from state $y$ is 
\begin{equation}
\bar \lambda^{\nu}_y \equiv \sum_{x\neq y} \bar k^{\nu}_{xy} = -\bar k^{\nu}_{yy}.
\end{equation}
We define a total rate matrix by $ \bar \bk \equiv \sum_\nu \bar \bk^{\nu}$ and a total escape-rate vector by $\bar \lambda \equiv \sum_\nu \bar \lambda^{\nu}$. These rates are functions of the affinities and dynamical bias; their analogs for the reference process are recovered at the point of vanishing of $a$ and $b$, namely $ \bk= \bar \bk (0,0)$ and $ \lambda = \bar \lambda(0,0)$. For the NE process, the state at time $\tau$ is $\bar z(\tau)$. A system state trajectory during time interval $[0,t]$ is denoted $[\bar z]$. The ensemble average over all trajectories $[\bar z]$ generated with dynamics corresponding to $ \bar \bk$ is $\l \dots \r_{\bar \bk} $. 

\subsection{Mapping typical NE fluctuations on rare EQ fluctuations \label{MappingNEonEQ}}

We now connect the energy currents, activities and occupations statistics for the EQ process with the statistics of the same variables for the stationary NE process. This mapping involves the \emph{escape-rate change} $c=c(a,b)$ defined by
\begin{equation}
c \equiv  \lambda - \bar \lambda, \label{EscapeRateChange}
\end{equation}
that is zero at vanishing affinities and dynamical biases. We emphasize that $c$ cannot be adjusted independently of $a$ and $b$. This means that the affinity and the dynamical bias are the central variables in determining the NESS reached by the system.

To connect EQ and NE fluctuations, one needs to redo all the calculations of sections \ref{EQ-LDF} and \ref{EQ-CGF}, but for the NE process, introducing the NE scaled CGF $\bar \Gamma = \bar \Gamma(\bar a, \bar b, \bar m) $ and LDF $\bar L=\bar L(j,f,p)$, the NE tilted operator $\bar \bkappa = \bar \bkappa(\bar  a, \bar b, \bar m)$, the generator of the NE driven process $\bar \bK = \bar \bK(\bar a, \bar b, \bar \u) $ and associated escape rate $\bar \Lambda=\bar \Lambda (\bar a, \bar b, \bar \u)$, the affinity increment $2\bar a $, the dynamical bias increment $\bar b$, the NE drift $\bar \u$ and the NE escape weight $\bar \m = \bar \m(\bar a, \bar b, \bar \u)$, all denoted with a bar to distinguish them from their equivalent for the EQ reference process. One obtains all these objects replacing $ k$ by $ \bar k$ and the Lagrange multipliers $(a,b,\u) $ by $ (\bar a,\bar b, \bar \u)$ in all the definitions. For instance, for the NE tilted operator, we have
\bea
\bar \kappa_{yy} &\equiv& -\sum_{x\neq y, \nu}  \bar k^{\nu}_{xy} + \bar \m_y, \label{NEDressedOperatorDiag}\\
\bar  \kappa_{xy} &\equiv&  \sum_{\nu}  \bar k^{\nu}_{xy} e^{\bar a_\nu(\e_x-\e_y)+\bar b_\nu d_{xy}}  .  \label{NEDressedOperator}
\eea
Notice that we call $2\bar a$ an affinity ``increment'' since we already deal with a NE process: a deviation from the typical current is associated with an ``increase'' of affinity that will make this fluctuation typical. For the same reason, the dynamical bias $ \bar b$ is also qualified as an increment.

The mapping between EQ and NE fluctuations now comes from the connection between the EQ and NE tilted operators 
\begin{equation}
\bar \bkappa(\bar a ,\bar b, \bar \m) = \bkappa(\bar a + a,\bar b +b,\bar \m +c), \label{DresOpmapppingNEtoEQ}
\end{equation}
that we obtain by comparing Eqs.~(\ref{DressedOperatorDiag}-\ref{DressedOperator}) with Eqs.~(\ref{NEDressedOperatorDiag}-\ref{NEDressedOperator}).
Hence, the same symmetry exists between the eigenvalues and between the eigenvectors: the \emph{full} spectrum of the two operators is connected. In particular, the scaled CGFs are connected by
\begin{equation}
\bar \Gamma(\bar a ,\bar b, \bar \m) = \Gamma(\bar a + a,\bar b +b,\bar \m +c), \label{CGFmappingNEtoEQ}
\end{equation}
and, from the Legendre transformation, the LDFs verify
\begin{equation}
\bar L(j,f,p) =  L(j,f,p)- a^\dag\cdot j - b^\dag\cdot f - c^\dag \cdot p. \label{LDFmappingNEtoEQ} 
\end{equation}
The left eigenvectors of the tilted operators satisfy
\begin{equation}
\bar \pi(\bar a ,\bar b, \bar \m)= \pi(\bar a + a,\bar b +b,\bar \m +c) \label{PimappingNEtoEQ}
\end{equation}
or equivalently
\begin{equation}
\bar \u(\bar a ,\bar b, \bar \m)= \u(\bar a + a,\bar b +b,\bar \m +c). \label{drifmappingNEtoEQ}
\end{equation}
The mapping also holds for the right eigenvectors and this leads to
\begin{equation}
\bar p(\bar a ,\bar b, \bar \m)= p(\bar a + a,\bar b +b,\bar \m +c). \label{ProbamappingNEtoEQ}
\end{equation}
Finally, the generators of the driven processes also verify
\begin{equation}
\bar \bK(\bar a, \bar b,\bar \u) = \bK(\bar a + a,\bar b +b, \u),
\end{equation}
where $\bar \u$ and $ \u$ are respectively the left- and right-hand sides of Eq.~(\ref{drifmappingNEtoEQ}).

Thus, the EQ and NE processes are tightly connected and one can focus on the EQ process' fluctuations only:
Eq.~(\ref{CGFmappingNEtoEQ}) shows that the statistics of energy currents, activities and occupations for any NE process with affinity $2a$ and dynamical bias $b$ is known from the statistics of the same variables computed for the EQ process. Indeed, the derivatives of Eq.~(\ref{CGFmappingNEtoEQ}) with respect to $\bar a$, $\bar b$ or $\bar \m$ evaluated in $(\bar a,\bar b,\bar \m) = (0,0,0)$ yields the NE cumulants of the energy currents, activities and occupations from the scaled CGF for the EQ reference process, e.g. for $j_\nu$ we have
\begin{equation}
\l j_\nu[\bar z] \r_{\bar \bk} =  \frac{\partial \bar \Gamma}{\partial \bar a_\nu}(0,0,0) = \frac{\partial \Gamma}{\partial a_\nu}(a,b,c).
\end{equation}
Notice that evaluating Eq.~(\ref{CGFmappingNEtoEQ}) at the point of vanishing of  $(\bar a ,\bar b, \bar \m)$ returns by definition of a scaled CGF
\begin{equation}
0 = \bar \Gamma(0,0,0) = \Gamma(a,b,c) , \label{AVGpoint}
\end{equation}
for all $a$ and $b$, with $c = (\lambda-\bar \lambda)$. Accordingly, the total derivatives of $ \Gamma(a,b,c)$ with respect to $a$ or $b$ also vanish exactly such  that $ \Gamma$ remains constant and equal to zero in the direction $(a,b,c)$. We call the subspace where $\Gamma$ vanishes the \emph{physical system subspace}: each point $(a,b,c)$ in this subspace defines a precise physical process with affinity $2a$ and dynamical bias $b$. The function $ \Gamma$ includes the full thermodynamic information on any system defined with the same energy levels $\e$, coupling matrices $\bgamma^{\nu}$ and energy barriers $\bm{d}$ (up to a reservoir specific dilatation), and so does the LDF $L$. One simply changes the degree of NE or the type of dynamics, encoded into the dilatation factors, by moving into the physical system subspace.

We end by remarking that the idea of mapping EQ and NE fluctuations was first proposed by Andrieux in Refs. \cite{ Andrieux2012_vola, Andrieux2012_vol}, but for the statistics of energy currents only. However, this mapping had no concrete application since the NE statistics of the currents were needed to define the EQ dynamics involved in the mapping. On the contrary, the mapping of Eq.~(\ref{CGFmappingNEtoEQ}) and (\ref{LDFmappingNEtoEQ}) is explicit, with the price that, when comparing with Ref. \cite{ Andrieux2012_vola, *Andrieux2012_vol}, the EQ statistics of activities and occupations must be known in addition to the energy currents statistics.

\subsection{Asymptotic equivalence of the driven reference process and the NE process \label{EnsembleEquivalence}}

We now discuss the asymptotic equivalence of the driven reference process and the NE process. We first prove the equality of their escape rates and on the way give a slightly simplified expression of $L$. Using this result, we demonstrate the equivalence of the path probabilities of the driven reference process and the NE process. 

From Eqs.~(\ref{exitrule}) and (\ref{AVGpoint}), we find $c + \Lambda -\lambda = 0$. This leads with Eq.~(\ref{EscapeRateChange}) to the equality of the escape rates of the driven reference process and the NE process
\begin{equation}
\Lambda= \bar \lambda, \label{DrivEQandNEescapeEquality}
\end{equation}
even though these two processes are different in general due to the drift $\u$, i.e. $K_{xy} \neq \bar k_{xy} $ if $x\neq y$. As a consequence, the LDF is written as
\begin{equation}
L(j,f,p) = a^\dag   \cdot j  + b^\dag \cdot f + c^\dag \cdot p.
\end{equation}
The equality of the escape rates indicates that the driven reference process and NE process look alike. Their generators are connected by the similarity transformation
\begin{equation}
\bK = |\bpi| \cdot \bar \bk \cdot | \bpi |^{-1},
\end{equation}
that follows from the comparison of Eqs.~(\ref{ExplicitBiasRate}) and (\ref{defNErates}). We denote $|\bpi|$ the positive and diagonal matrix obtained by taking the absolute value of the elements of $\bpi$. The equality of the diagonal part of the Markov matrices of the two processes is granted by Eq.(\ref{DrivEQandNEescapeEquality}). From this similarity transformation, one can show the asymptotic equality of the path probabilities associated to each process
\begin{equation}
\P_{\bK}[y] \underset{t \rightarrow \infty}{\simeq} \P_{\bar \bk}[y],
\end{equation}
for any trajectory $[y]$. We have defined the path probabilities knowing the initial state $y(0)$
\begin{equation}
\P_{\bar \bk}[y] \equiv  \exp \left(  -\int_0^t \D \tau \bar \lambda_{y(\tau)} \right) \prod_{\tau\in [0,t]}\bar k^{\nu(\tau)}_{y(\tau+d\tau)y(\tau)},
\end{equation}
for the NE process and 
\begin{equation}
\P_{\bK}[y] \equiv  \exp \left(  -\int_0^t \D \tau \Lambda_{y(\tau)} \right) \prod_{\tau\in [0,t]} K^{\nu(\tau)}_{y(\tau+d\tau)y(\tau)},
\end{equation}
for the driven reference process.
In these equations, the product applies for all times $\tau$ at which the system changes of state during the trajectory $[y]$, with $y(\tau)$ (respectively $y(\tau+d\tau)$) the system state before (respectively after) the transition at time $\tau$. The exponential terms appearing in these two path probabilities are equal. Concerning the product terms, they differ from boundary terms only
\bea
\prod_{\tau} K^{\nu(\tau)}_{y(\tau+d\tau)y(\tau)} &=& \prod_{\tau} |\pi_{y(\tau+d\tau)}| \bar k^{\nu(\tau)}_{y(\tau+d\tau)y(\tau)} |\pi_{y(\tau)}| ^{-1}, \nonumber \\
&=& |\pi_{y(t)}| \left( \prod_{\tau} \bar k^{\nu(\tau)}_{y(\tau+d\tau)y(\tau)} \right) |\pi_{y(0)}| ^{-1}.  \nonumber \\
\eea
Then, the path probabilities of the driven reference process and NE process verify 
\begin{equation}
\lim_{t\rightarrow \infty} \frac{1}{t} \ln \frac{\P_{\bar \bk} [y]}{\P_{\bK} [y]} =0,
\end{equation}
and are asymptotically equivalent \cite{Chetrite2015_vol16}. Since the driven reference process is the dynamics that typically reproduces the conditioned reference process, we conclude that there is an ensemble equivalence between the NE process and the conditioned reference process. This central result is similar to the path-ensemble equivalence derived in Refs. \cite{Chetrite2015_vol16,Chetrite2013_vol111}. In Appendix \ref{EnsembleEquivalence2}, we show that the NE process is asymptotically equivalent to the canonical process that is defined by exponentially weighting each trajectory, even though these two processes are not exactly identical.

\section{Discussion and general summary \label{discussion}}

%%%%%%%%%%%%%%%%%%%%%%%%%%%%%%%%%%%%%%%
\begin{table*}
	\caption{EQ and NE thermodynamic potentials.  \label{summary}}
	\begin{tabular}{lcccc}
		\hline 
		\hline
		Ensemble 	&  \quad Micro-canonical \quad		& \quad Canonical \quad	& \quad NE micro-canonical \quad 		& \quad  Meta-canonical \quad			\\
		\hline
		Potential 				&  $s = -\sum_x p_x \ln p_x$ 		&  $\varphi = -\ln \l \exp{(-\beta_1 e[z])} \r_{\bk}$  	& $L(j,f,p)$ 			&  $\Gamma (a,b,\m)$  	\\
		Variational principle   	& max    		&  min						& max    				& max  \\
		Free variables 			& $  \beta_1$ 	& $e$ 						& $ a  $, $ b $, $\m$   &  $ j $, $ f $, $p$ \\
		Constrained variables 	&	$ e $ 		& $   \beta_1 $ 			& $ j $, $ f $ , $p$			& $ a $, $ b $, $\m$  \\
		Physical system subspace&--- 			&---			 			& $\m(j,f,p) = c\big(a(j,f,p),b(j,f,p)\big)$  & $\m =c(a,b)$ \\
		No dilatation space		&--- 			&---			 			& $b(j,f,p) = 2a(j,f,p)$	& $b=2a$ \\
		Legendre structure		& \multicolumn{2}{c}{$s + \varphi = \beta_1 e $  }
								& \multicolumn{2}{c}{$L + \Gamma = a^\dag \cdot  j + b^\dag \cdot f + \m^\dag \cdot p $ }\\
		\hline 
		\hline
	\end{tabular}
\end{table*}
%%%%%%%%%%%%%%%%%%%%%%%%%%%%%%%%%%%%%%%

In Sec.\ref{EQfluctuation}, we have studied the fluctuations of an EQ system exchanging energy with several heat reservoirs at the same temperature. We have seen that energy may spontaneously flow from one reservoir to another, even if it does not on average. Each of these current fluctuations has been associated to a temperature difference that would typically reproduce it. Similarly, we have shown that a fluctuation of the activity of the exchanges with each reservoir would be typically reproduced by dilating the appropriated energy barriers. From these observations, we have identified two couples of conjugated variables and provided the corresponding LDF and CGF from a variational approach. 

In Sec.\ref{NEfluctuations}, we have considered the fluctuations of the system defined in Sec.\ref{EQfluctuation}, but driven out-of-equilibrium by temperature differences between the heat reservoirs. We have found an exact mapping between the statistics of the energy currents, activities and occupations for the EQ and NE systems. We have also discussed the asymptotic equivalence of the trajectory ensembles generated by the conditioned EQ process and the NE process. From the existence of the mapping between EQ and NE systems, we have concluded that the study of a NE system amounts to the calculation of the probability of rare fluctuations of the same system at EQ. Now that the distinction between the dynamical fluctuations of EQ and NE systems has been dispelled, we come back to the results of Sec.\ref{EQfluctuation} and summarize the canonical structure satisfied by the two ensembles of NE systems.

\subsection{Summary of the NE canonical structure}
\label{CanonicalStructure}

The ensemble of systems in contact with several heat reservoirs at different temperatures is called the \emph{meta-canonical ensemble}. The trajectories of the systems in the meta-canonical ensemble are generated by the NE process with generator $\bar \bk$.
All the systems in this ensemble have the same energy levels $\e_x$, and the same dynamical parameters, i.e. energy barriers $d_{xy}$ and couplings  with the heat reservoirs $\gamma^\nu_{xy}$. By convention, the heat reservoir of smallest temperature is the reference reservoir ($\nu=1$) such that all the affinities $2a_\nu = \beta_1 - \beta_\nu$ are positive. Notice that the temperature of the reference reservoir sets the energy scale and has no physical relevance. On the opposite, the affinities $a_\nu$ are the central variables of the meta-canonical ensemble that are set by the environmental constraints. The affinities are naturally conjugated to the energy currents. However, we know from the previous sections that considering $(a,j)$ as the unique couple of conjugated variables does not afford to study all NE systems from the same NE potential. Intuitively, a change of an affinity also impacts the system activity and the occupation of the various states. Hence, we have introduced additional intensive variables to take into account these effects separately: the dynamical biases connected to the dilatation factors of the energy barriers and the escape weights modifying the escape probability of each state. These two intensive variables cannot be adjusted independently of the affinities if we want to avoid a change of the system dynamics: no dilatation should be applied to the energy barriers ($l_\nu =1$ for all $\nu$) yielding to dynamical biases that are equal to the affinities ($b = 2 a $); the dynamics should conserve the norm of the occupation vector imposing that an affinity must be associated with an escape weight equal to the escape-rate change $\m=c(a,2a)$. Therefore, in the meta-canonical ensemble, the environment sets the affinity vector $a$ which in turn constrains the dynamical intensive variables, namely the dynamical biases and the escape weights. The NE potential of the meta-canonical ensemble is the CGF of energy currents, activities and occupations $\Gamma(a,b,\m)$. It vanishes for all $a$ when $b=2a$ and $\m=c(a,2a)$, but its partial derivatives with respect to $a$, $b$ and $\m$ produces all the NESS cumulants of energy currents, activities and occupations for any affinity. For instance, the thermodynamic behavior follows from the NE equations of state
\begin{align}
  \left. \frac{\partial \Gamma}{ \partial a_\nu } \right|_{a_{\smallsetminus \nu},b,\m } &= j_\nu, \label{CanonicalRelNESS1} \\
  \left. \frac{\partial \Gamma}{ \partial b_\nu } \right|_{a,b_{\smallsetminus \nu},\m } &= f_\nu, \label{CanonicalRelNESS2}     \\ 
  \left. \frac{\partial \Gamma}{ \partial \m_x } \right|_{a,b,\m_{\smallsetminus x} } &= p_x,      \label{CanonicalRelNESS3}
\end{align}
where the subscripts on the vertical bars indicate variables that remain constant when taking the partial derivative. We denote $ a_{\smallsetminus \nu} $ the vector $ a $ without the $ \nu $\textit{th} component. The cumulants of EQ thermodynamic variables are obtained with the NESS occupations defined by $p^*=p(a,2a,c)$ that only depend on $\chi-1$ affinities. The mean energy in the NESS is $ \l e[\bar z] \r_{\bar \bk}  = e( p^*) $, and the mean entropy is $ \l s[\bar z] \r_{\bar \bk} =s(p^*)$. 

The ensemble of systems conditioned on the energy currents they received from their environment is called the \emph{NE micro-canonical ensemble}. The trajectories of the systems in this ensemble are generated by the EQ reference process with generator $\bk$ filtrated to achieve the condition on the energy currents. The physical implementation of systems in the NE micro-canonical ensemble would require the existence of energy sources with no fluctuations. These sources will very likely not exist in practice \bibnote{Beyond the case of energy sources, one can imagine an experimental setup where a matter flux or a rate of volume increase can be controlled exactly.}, even though this problem is not specific to NE ensembles (see for instance page 83 of Ref.~\cite{Callen1985_vol} for an example in EQ thermodynamic theory). If we assume that an energy current can be imposed from the outside, the activities and the occupations must take precise values so that the system can sustain the energy current.
On the opposite, the conjugated intensive variables become free to fluctuate. The relationship between currents, activities and occupations is obtained from the correspondence between the conjugated variables $(j,f,p)$ and $(a,b,\m)$, as summarized in Table \ref{summary}. The NE micro-canonical potential is the LDF $L(j,f,p)$ and the statistics of the intensive variables $(a,b,\m)$ follows from its partial derivative
\begin{align}
 \left. \frac{\partial L }{ \partial j_\nu } \right|_{ j_{\smallsetminus \nu},f,p} &= a_\nu,  \label{microCanonicalRelNESS1} \\
 \left. \frac{\partial L }{ \partial f_\nu } \right|_{ j,f_{\smallsetminus \nu},p } &= b_\nu, \label{microCanonicalRelNESS2}     \\ 
 \left. \frac{\partial L }{ \partial p_x } \right|_{ j,f,p_{\smallsetminus x} } &= \m_x.      \label{microCanonicalRelNESS3}
\end{align}

We proved in sections \ref{EQfluctuation} and \ref{EnsembleEquivalence} the equivalence of the ensembles of trajectories generated by the NE process and the conditioned EQ reference process assuming that the NE potentials are convex. Accordingly, the meta-canonical ensemble and NE micro-canonical ensembles are ensemble equivalent. In other words, systems submitted to temperature gradients are equivalent, at the thermodynamic level, to systems subjected to stationary energy injection (and extraction). By construction, the NE potentials are conjugated by Legendre transformation
\bea
L(j,f,p) + \Gamma(a,b,\m) &=& a^\dag \cdot j + b^\dag \cdot f + \m^\dag \cdot p, \label{LegL}
\eea
and the NE stationary state can be obtained from a variational approach.
If we consider $ a^\dag \cdot j + b^\dag \cdot f + \m^\dag \cdot p  - \Gamma(a,b,\m) $ as the potential $ L $ that would be obtained from Eq.~(\ref{LegL}) by assuming the independence of the conjugated variables, then the NESS affinity, dynamical bias and escape weight reached by the system at constant imposed energy current $ j $, activity $f$, and occupation $p$ maximize this potential in the subspace of constant $ (j,f,p) $:
\begin{equation}
(a,b,\m)= \underset{a,b,\m| j,f,p }{\arg\!\max} \left[ a^\dag \cdot j + b^\dag \cdot f + \m^\dag \cdot p - \Gamma(a,b,\m) \right]
\end{equation}
which are exactly Eqs.~(\ref{CanonicalRelNESS1}-\ref{CanonicalRelNESS3}). 
The same argument holds the other way around.
If we consider $ a^\dag \cdot j  +b^\dag \cdot f + \m^\dag \cdot p  - L(j,f,p) $ as the potential $ \Gamma $ that would be obtained from Eq.~(\ref{LegL}) assuming  the independence of the conjugated variables, then the NESS energy currents, activities and occupations reached by the system at constant imposed affinity $ a $, dynamical bias $b$, and escape weight $\m$ maximize this potential in the subspace of constant $ (a,b,\m) $:
\begin{equation}
(j,f,p) = \underset{j,f,p | a,b,\m }{\arg\!\max} \left[ a^\dag \cdot j + b^\dag \cdot f + \m^\dag \cdot p  - L(j,f,p) \right]
\end{equation}
which are exactly Eqs.~(\ref{microCanonicalRelNESS1}-\ref{microCanonicalRelNESS3}). 

\subsection{Symmetries of the NE potentials \label{NEsymmetries}}

The metacanonical potential is even under the sign change of all affinities. We prove in Appendix \ref{FT} that this symmetry leads to the fluctuation theorem (FT), a fundamental result regarding the asymptotic statistics of entropy production first studied in Refs. \cite{Bochkov1981_vol106a, Evans1994_vol50, Gallavotti1995_vol74}. Another fundamental symmetry is obtained from the equality of second derivatives of the NE potentials. This symmetry is the NE equivalent of the Maxwell relations and reads as
\begin{equation}
\frac{\partial^2 \Gamma}{ \partial h_{\alpha} \partial h'_{\alpha'} } = \frac{\partial^2 \Gamma}{ \partial h'_{\alpha'} \partial h_{\alpha} } \quad \mbox{and} \quad
\frac{\partial^2 L}{ \partial v_{\alpha} \partial v'_{\alpha'} } = \frac{\partial^2 L}{ \partial v'_{\alpha'} \partial v_{\alpha} }
\label{MaxwellRelations}
\end{equation}
where $h$ and $h'$ are two vectors in $(a,b,\m)$ and similarly $v$ and $v'$ in $(j,f,p)$. 
The subscripts $\alpha$ and $\alpha'$ indicate two arbitrary components of these vectors.
At EQ,  Maxwell's relations deeply constrain the number of EQ response coefficients that should be introduced to completely describe a system. Here, they constrain the derivatives of the non-linear functions giving, for instance, the currents in terms of the affinities. In the close-to-EQ limit, Eq.~(\ref{MaxwellRelations}) implies that the linear response matrix is symmetric, or in other words it implies the Onsager reciprocity relations \cite{Onsager1931_vol37,Andrieux2004_vol121}, as we will see in the next section. 

\subsection{NE linear response theory \label{NEresponse}}

We study the linear response of a system in an arbitrary NESS and further perturbed by a change of temperature $\beta_\nu \rightarrow \beta_\nu' = \beta_\nu + \Delta \beta_\nu $ or of dilatation factor $l_\nu \rightarrow l_\nu'= l_\nu + \Delta l_\nu$. More precisely, we want to determine the change of the energy currents and activities when the half affinities $ a_\nu = (\beta_1 - \beta_\nu)/2 $ and dynamical biases $ b_\nu = (\beta_1 l_1 - \beta_\nu l_\nu) $ are slightly changed to the new values $   a_\nu + \Delta a_\nu $ and $  b_\nu + \Delta b_\nu$. We assume that $ l_1 $ and $ \beta_1 $ do not change during the perturbation. Then, the perturbations are written as
\begin{eqnarray}
  \Delta a_\nu &=& - \Delta \beta_\nu/2, \\
  \Delta b_\nu &=& \left( - \beta_\nu + 2\Delta a_\nu \right) \Delta l_\nu  + 2 l_\nu \Delta a_\nu \nonumber \\ 
               & \simeq &  -\beta_\nu \Delta l_\nu  +2 l_\nu \Delta a_\nu, 
\end{eqnarray}
at linear order. We remark that the dynamical biases change when perturbing the affinities, but the converse is not true. 

A Taylor expansion of the meta-canonical potential $  \Gamma $ gives the following quadratic function 
\begin{multline}
\Gamma( a+\Delta a, b+\Delta b,  \m+\Delta \m)  = \\
 \Gamma( a,  b,  \m) + \sum_{h=a,b,\m} \Delta h^\dag \cdot \nabla_h \Gamma  \\
 + \frac{1}{2}\underset{h'=a,b,\m}{\sum_{h\;=a,b,\m}} \Delta h^\dag \cdot \nabla_{hh'} \Gamma \cdot \Delta h', \label{QuadraticApproxGamma}
\end{multline}
where $\Delta \m$ is not yet specified. We have used the short notations for the derivatives of the meta-canonical potential
\bea
\left( \nabla_h \Gamma \right)_\alpha  &\equiv& \frac{\partial \Gamma}{\partial h_\alpha }\left( a,  b,  \m \right) , \\
\left( \nabla_{hh'} \Gamma \right)_{\alpha\alpha'} &\equiv& \frac{\partial^2 \Gamma}{\partial h_\alpha \partial h'_{\alpha'}} \left( a, b,  \m \right).
\eea
The perturbation induces a variation $\Delta j$ of the energy currents, $\Delta f$ of the activities or $\Delta p$ of the occupation. Taking the partial derivative of Eq.~(\ref{QuadraticApproxGamma}) with respect to $ \Delta a $, $ \Delta b $, or $ \Delta \m$ and evaluated in $\Delta \m = \Delta c $, with $\Delta c$ the variation of the escape-rate change due to the perturbation, leads to the linear response equation
\begin{equation} \left(
\begin{array}{c}
\Delta j \\ 
\Delta f \\
\Delta p
\end{array} \right)
 \simeq  \left[
\begin{array}{ccc} 
     \nabla_{aa} \Gamma  	&    \nabla_{ab} \Gamma  	&    \nabla_{a\m} \Gamma \\
     \nabla_{ba} \Gamma  	&    \nabla_{bb} \Gamma  	&    \nabla_{b\m} \Gamma \\
     \nabla_{\m a} \Gamma &    \nabla_{\m b} \Gamma &    \nabla_{\m\m} \Gamma \\
\end{array} \right]
\cdot \left(
\begin{array}{c}
	\Delta a \\ 
	\Delta b \\
	\Delta c
\end{array} \right). \label{NELinearResponse}
\end{equation}
From Eq.~(\ref{MaxwellRelations}), the response matrix above is symmetric even close to an arbitrary NESS. However, the chain rule yields
\begin{equation}
\Delta c =   \nabla_{a}c \cdot \Delta a  +   \nabla_{b}c \cdot \Delta b,
\end{equation}
and the variation of the currents and activities becomes
\begin{align}
\Delta j =& \left(  \nabla_{aa} \Gamma +  \nabla_{a \m} \Gamma \cdot  \nabla_{a}c \right) \cdot \Delta a \nonumber \\ & \qquad\qquad + \left(  \nabla_{ab} \Gamma +  \nabla_{a \m} \Gamma \cdot  \nabla_{b}c \right) \cdot \Delta b. \label{CurrResp} \\
\Delta f =& \left(  \nabla_{ab} \Gamma +  \nabla_{b\m} \Gamma \cdot  \nabla_{a}c \right) \cdot \Delta a \nonumber \\ & \qquad\qquad + \left(  \nabla_{bb} \Gamma +  \nabla_{b\m} \Gamma \cdot  \nabla_{b}c \right) \cdot \Delta b. \label{FrenResp}
\end{align}
The response matrix defined from Eqs.~(\ref{CurrResp}-\ref{FrenResp}) is no longer symmetric in general as already emphasized in former works on NE linear response theory \cite{Baiesi2009_vol137, Baiesi2009_vol103,Verley2011_vol93,Verley2011_vol10, Speck2006_vol74, Chetrite2008_vol2008, Chetrite2009_vol137, Seifert2010_vol89}. 
The second derivatives of the meta-canonical potential appearing in Eq.~(\ref{CurrResp}) are
\begin{align}
\left (  \nabla_{aa} \Gamma\right)_{\nu \nu'} &= \lim_{t\rightarrow \infty} t \left\{ \l j_\nu[\bar z] j_{\nu'}[\bar z] \r_{\bar \bk} -  \l j_\nu[\bar z] \r_{\bar \bk} \l j_{\nu'}[\bar z] \r_{\bar \bk} \right\} , \nonumber \\
\left (  \nabla_{ab} \Gamma\right)_{\nu \nu'} &= \lim_{t\rightarrow \infty} t \left\{ \l j_\nu[\bar z] f_{\nu'}[\bar z] \r_{\bar \bk} -  \l j_\nu[\bar z] \r_{\bar \bk} \l f_{\nu'}[\bar z] \r_{\bar \bk}\right\} , \nonumber \\
\left (  \nabla_{a\m} \Gamma\right)_{\nu x}  &= \lim_{t\rightarrow \infty} t \left\{ \l j_\nu[\bar z] p_{x}[\bar z] \r_{\bar \bk} -  \l j_\nu[\bar z] \r_{\bar \bk} \l p_{x}[\bar z] \r_{\bar \bk}\right\} , \label{2ndeDerivGamma}
\end{align}
and correspond respectively to the current-current, the current-activity and the current-occupation covariances in the unperturbed NESS \bibnote[time]{We have multiplied the covariances by a factor $t$ since we consider time-averaged variables; we would have divided by $t$ if we were considering the time-integrated variables.}. In addition to the above covariances, the response functions include another contribution involving the derivatives of the escape-rate change $c$. Since the escape-rate change satisfies
\bea
- \frac{\partial c_x}{\partial a_\nu} &=& \sum_y \bar k^{\nu}_{yx} (\e_y - \e_x), 
\eea
the unperturbed mean-occupation multiplied by this derivative returns the unperturbed mean energy current
\begin{equation}
- \sum_{x}\frac{\partial c_x}{\partial a_\nu} \l p_{x}[\bar z] \r_{\bar \bk}  =  \sum_{x,y}  \bar k^{\nu}_{yx} \l p_{x}[\bar z] \r_{\bar \bk} (\e_y - \e_x) =  \l j_\nu[\bar z] \r_{\bar \bk} .
\end{equation}
Therefore, the response to the affinity perturbation is 
\begin{multline}
\left( \nabla_{aa} \Gamma +  \nabla_{a \m} \Gamma \cdot  \nabla_{a}c \right)_{\nu\nu'}  \\
= \lim_{t \rightarrow \infty} t \left\{ \l \vphantom{\frac{1}{1}} j_\nu[\bar z] j_{\nu'}[\bar z] \r_{\bar \bk} -  \l j_\nu[\bar z] \frac{\partial }{\partial a_{\nu'}} \left( p^\dag[\bar z] \cdot \bar \lambda \right) \r_{\bar \bk}\right\}. \label{AffResponse}
\end{multline}
As expected, the response has an additive structure with an equilibriumlike part given by a currents correlation function, and a NE part corresponding to a current and traffic-excess correlation function. We call traffic-excess the derivative of the empirical escape rate $p^\dag[\bar z] \cdot \bar \lambda $ with respect to the perturbed variable \cite{Baiesi2009_vol137, Baiesi2009_vol103}. Similarly, the response of the energy current to a perturbation of the dynamical bias in the second line of Eq.~(\ref{CurrResp}) has two parts with an activity-current correlation function and a current-traffic excess correlation function. 

As regards the perturbation of an EQ system, i.e. all $\beta'_\nu $ are close to the reference inverse temperature $ \beta_1$, one recovers the Yamamoto--Zwanzig formula expressing the response coefficients to a temperature perturbation from the covariances of energy currents \cite{Yamamoto1960_vol33,Zwanzig1965_vol16}. In order to see this, let us first consider a reference system at EQ only perturbed by a change of the dilatation factors, i.e. $ \Delta a = 0$ and $ \Delta b = -\beta_1 \Delta l  $. Thanks to Eq.~(\ref{CurrResp}), the variation of the energy currents is written as
\begin{equation}
\Delta j = \left( \nabla_{ab} \Gamma + \nabla_{a\m} \Gamma \cdot \nabla_{b}c \right) \cdot \Delta b  = 0.
\end{equation}
It vanishes for any perturbations $ \Delta b $ since no mean energy current exists at EQ. Thus, we find
\begin{equation}
\nabla_{ab} \Gamma + \nabla_{a\m} \Gamma \cdot \nabla_{b}c = 0,
\end{equation}
if the derivatives are taken in $a=0$. This removes the contribution due to the dynamical bias from the EQ response. Another contribution disappears in the close-to-equilibrium limit due to the decoupling between occupations and energy currents \cite{Wynants2010_vola}. Indeed, from the symmetry of the meta-canonical potential with sign change of the affinities, namely $\Gamma(a,b,\m) = \Gamma(-a,b,\m)$, we have
\beq
\frac{\partial^2 \Gamma}{ \partial a_{\nu} \partial \m_x }(a,b,\m) = - \frac{\partial^2 \Gamma}{ \partial a_{\nu} \partial m_x } (-a,b,\m).
\eeq
Accordingly, $\nabla_{a\m} \Gamma = 0$ if the derivatives are taken in $a=0$. From the third line of Eq.~(\ref{2ndeDerivGamma}), we can conclude that the energy currents and occupations are decoupled. The Yamamoto--Zwanzig formula follows from Eq.~(\ref{CurrResp})
\beq
\Delta j = \frac{\nabla_{aa} \Gamma}{2} \cdot (\beta_1 - \beta'),
\eeq
where $\nabla_{aa} \Gamma$ is given in the first line of Eq.~(\ref{2ndeDerivGamma}) with EQ averages $\l \cdots \r_{\bk}$ instead of the NE averages $\l \cdots \r_{\bar \bk}$. Therefore, we recover the Onsager reciprocity relations from the NE Maxwell-relations.

\section{Illustrative example: a two-level system \label{Example}}

We now illustrate our results on a two-level system with states $z=1,2$ and mechanisms $\nu=1,2, \cdots, \chi$ enabling energy exchanges with $\chi$ different heat reservoirs. The coupling strength with reservoir $\nu$ is denoted $\gamma_\nu$ in this section since it is not a matrix but a vector when there are only two states. The energy states are $\e_1$ and $ \e_2$. Let $\e_\pm = \e_1\pm \e_2$ to shorten notations. The transition rate matrix of the EQ reference process for each mechanism $\nu$ is
\begin{equation}
  \bk^{\nu} = \left[ \begin{matrix}
-\gamma_\nu e^{  \frac{\beta_1 \e_-}{2}} & 
 \gamma_\nu e^{ -\frac{\beta_1 \e_-}{2}} \\
 \gamma_\nu e^{  \frac{\beta_1 \e_-}{2}} & 
-\gamma_\nu e^{ -\frac{\beta_1 \e_-}{2}} 
\end{matrix} \right] \label{ExRefProcess}
\end{equation}
where we assume vanishing dilatation factors $l_1$, see Eq.~(\ref{defEQrates}). 
The rate matrices for the NE system are 
\begin{equation}
 \bar \bk^{\nu} = \left[ \begin{matrix}
-\gamma_\nu e^{  (\beta_1/2 - a_\nu) \e_- + b_\nu \e_+} & 
 \gamma_\nu e^{ -(\beta_1/2 - a_\nu) \e_- + b_\nu \e_+} \\
 \gamma_\nu e^{  (\beta_1/2 - a_\nu) \e_-  + b_\nu \e_+} & 
-\gamma_\nu e^{ -(\beta_1/2 - a_\nu) \e_- + b_\nu \e_+} 
\end{matrix} \right], \label{ExNEProcess}
\end{equation}
if we chose $d_{12}= \e_+$.
The escape-rate changes for this model are
\bea
c_1 &=& \sum_\nu \gamma_\nu e^{ \beta_1 \e_-/2} \left( 1-e^{ - a_\nu \e_- + b_\nu \e_+}  \right), \label{EscChange} \\
c_2 &=& \sum_\nu \gamma_\nu e^{-\beta_1 \e_-/2} \left( 1-e^{   a_\nu \e_- + b_\nu \e_+}  \right).
\eea
The tilted operator $ \bkappa =  \bkappa(a,b,\m)$ for the EQ reference system is 
\begin{equation}
  \bkappa= \! \left[ \begin{matrix} \displaystyle
-\sum_\nu\gamma_\nu e^{  \frac{\beta_1 \e_-}{2}} +\m_1  &  \displaystyle
  \sum_\nu \gamma_\nu e^{-(\beta_1/2 - a_\nu) \e_- +b_\nu \e_+}  \\
 \displaystyle \sum_\nu \gamma_\nu e^{ (\beta_1/2 - a_\nu) \e_- +b_\nu \e_+} & \displaystyle
-\sum_\nu \gamma_\nu e^{ -\frac{\beta_1 \e_-}{2}} +\m_2
\end{matrix} \right].
\end{equation}
The highest eigenvalue of this matrix is the meta-canonical potential
 \begin{multline}
\Gamma = -\sum_\nu \gamma_\nu \cosh \left(\beta_1 \e_-/2 \right)+\frac{\m_1+\m_2}{2} \\
+  \sqrt{ \hat \gamma^2+ \sum_{\nu,\nu'} \gamma_\nu \gamma_{\nu'} e^{(a_\nu- a_{\nu'} )\e_- + (b_\nu + b_{\nu'})\e_+}},
\end{multline}
where we have introduced
\begin{equation}
\hat \gamma \equiv -\sum_\nu \gamma_\nu\sinh \left( \frac{\beta_1 \e_-}{2} \right)+\frac{\m_1-\m_2}{2}. \label{hatgamma}
\end{equation}
The meta-canonical potential $\Gamma$ provides the statistics of $j_\alpha $ the energy current flowing from the $\alpha$th reservoir toward the system and of $f_\alpha$ the activity induced by the $\alpha$th mechanism. From direct derivation of $\Gamma$ with respect to $a_{\alpha}$, $b_{\alpha}$ or $\m_z$ the energy current coming from reservoir $\alpha > 1 $ is
\begin{equation}
j_\alpha = \frac{\sum_\nu \e_-\gamma_\alpha\gamma_\nu e^{(b_\nu+b_\alpha) \e_+}  \sinh \left[  ( a_\alpha -  a_\nu) \e_- \right]}{\sqrt{\hat \gamma^2+ \sum_{\nu,\nu'} \gamma_\nu \gamma_{\nu'} e^{( a_\nu-  a_{\nu'} )\e_- + (b_\nu + b_{\nu'})\e_+} }}, \label{meanj}
\end{equation}
\\*
the activity for the transitions induced by mechanism $\alpha$ is
\begin{equation}
f_\alpha = \frac{\sum_{\nu} \e_+\gamma_\alpha \gamma_\nu e^{(b_\nu + b_\alpha) \e_+}  \cosh\left[ ( a_\alpha -  a_\nu) \e_-  \right]}{\sqrt{\hat \gamma^2+ \sum_{\nu,\nu'} \gamma_\nu \gamma_{\nu'} e^{( a_\nu-  a_{\nu'} )\e_-/ + (b_\nu + b_{\nu'})\e_+} }},
\label{meanf}
\end{equation}
and the occupation of state $z$ is
\begin{equation}
p_z = \frac{1}{2}+ \frac{(\delta_{z,1}-\delta_{z,2}) \hat \gamma /2 }{\sqrt{ \hat \gamma ^2 + \sum_{\nu,\nu'} \gamma_\nu \gamma_{\nu'} e^{(a_\nu- a_{\nu'} )\e_- + (b_\nu + b_{\nu'})\e_+}}} \label{NESSprob}
\end{equation}
where $\hat \gamma$ of Eq.~(\ref{hatgamma}) is evaluated in $\m = c(a,2a)$, and taking $b=2a$ to obtain the mean values of $j$, $f$ and $p$ in the NESS with affinity $2a$. 
Deriving once more with respect to $ a_{\alpha'}$, $b_{\alpha'}$ or $\m_{z'}$ leads to the symmetric response matrix, see Eq.~(\ref{NELinearResponse}). 
The left and right eigenvectors of $\bkappa$ associated to the eigenvalue $\Gamma $ are respectively $\pi$ and $r=\bpi^{-1} \cdot p$. We find for the two-level model
\begin{widetext}
\bea
\pi_1 &=& \frac{ \sum_\nu \gamma_\nu e^{ (\beta_1/2-a_\nu) \e_- +b_\nu \e_+}}{\sum_\nu \gamma_\nu e^{ (\beta_1/2-a_\nu) \e_- +b_\nu \e_+} +\sum_\nu \gamma_\nu e^{  \beta_1 \e_-/2 }-\m_1  +\Gamma },  \label{pi1}\\
\pi_2 &=& \frac{ \sum_\nu \gamma_\nu e^{  \beta_1 \e_-/2 }-\m_1  +\Gamma }{\sum_\nu \gamma_\nu e^{ (\beta_1/2-a_\nu) \e_- +b_\nu \e_+} +\sum_\nu \gamma_\nu e^{  \beta_1 \e_-/2 }-\m_1  +\Gamma },   \label{pi2} \\
r_1 &=& \frac{ \left( \sum_\nu \gamma_\nu e^{ -(\beta_1/2-a_\nu) \e_- +b_\nu \e_+} \right)  \left( \sum_\nu \gamma_\nu e^{ (\beta_1/2-a_\nu) \e_- +b_\nu \e_+} +\sum_\nu \gamma_\nu e^{  \beta_1 \e_-/2 }-\m_1  +\Gamma \right) }{\sum_{\nu,\nu'} \gamma_\nu \gamma_{\nu'} e^{(a_\nu- a_{\nu'} )\e_- + (b_\nu + b_{\nu'})\e_+} +\left(\sum_\nu \gamma_\nu e^{  \beta_1 \e_-/2 }-\m_1  +\Gamma \right)^2}, \\
r_2 &=& \frac{ \left( \sum_\nu \gamma_\nu e^{  \beta_1 \e_-/2 }-\m_1  +\Gamma \right)  \left( \sum_\nu \gamma_\nu e^{ (\beta_1/2-a_\nu) \e_- +b_\nu \e_+} +\sum_\nu \gamma_\nu e^{  \beta_1 \e_-/2 }-\m_1  +\Gamma \right) }{\sum_{\nu,\nu'} \gamma_\nu \gamma_{\nu'} e^{(a_\nu- a_{\nu'} )\e_- + (b_\nu + b_{\nu'})\e_+} +\left(\sum_\nu \gamma_\nu e^{  \beta_1 \e_-/2 }-\m_1  +\Gamma \right)^2}. \label{rightEigVec}
\eea
\end{widetext}
%%%%%%%%%%%%%
\begin{figure*}
\begin{center}
\begin{tabular}{lll}
\includegraphics[width=6cm]{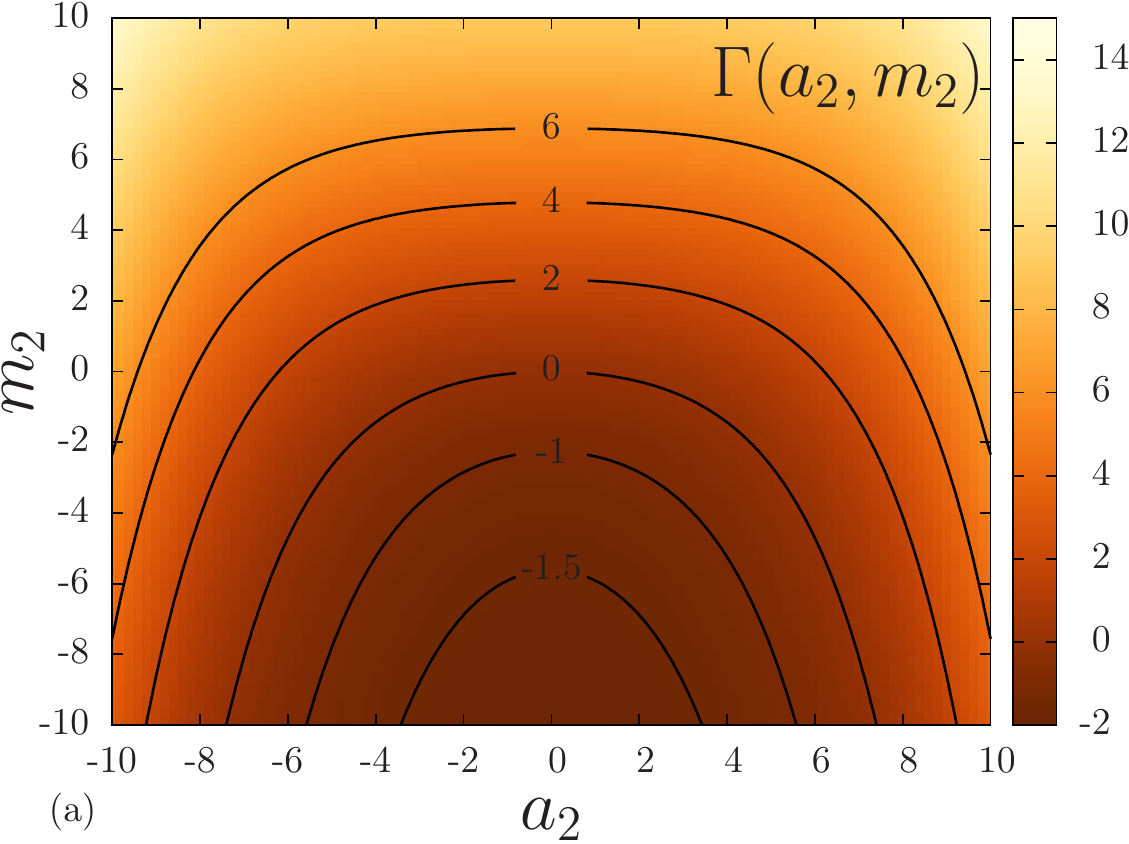} 
& \includegraphics[width=6cm]{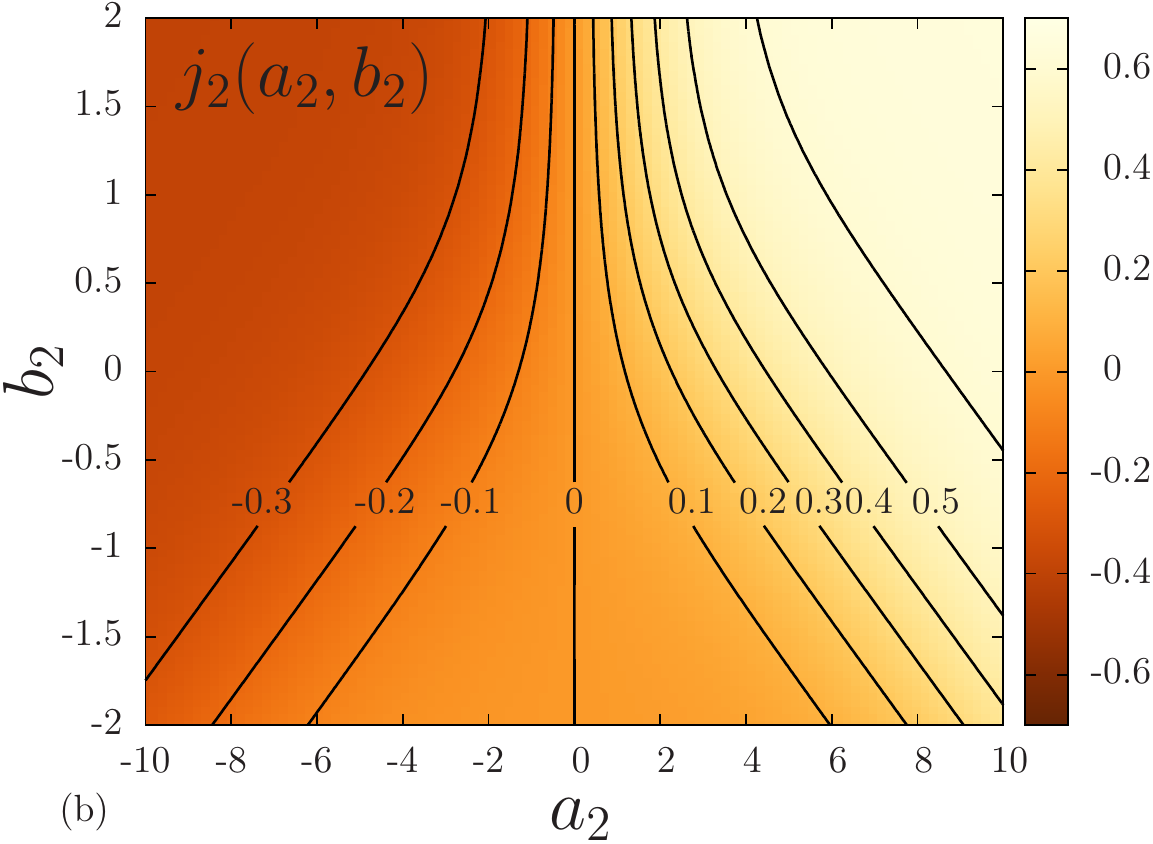} & \includegraphics[width=6cm]{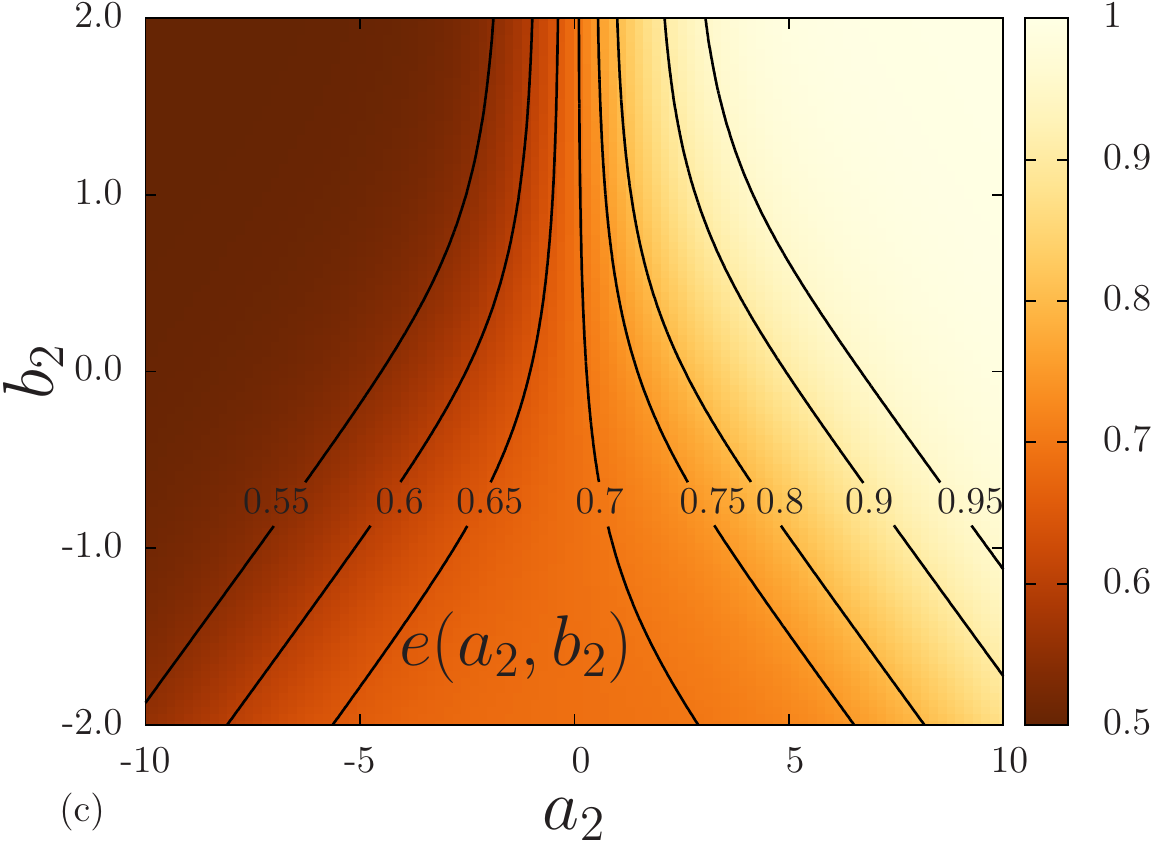}  \\  && \\
\includegraphics[width=6cm]{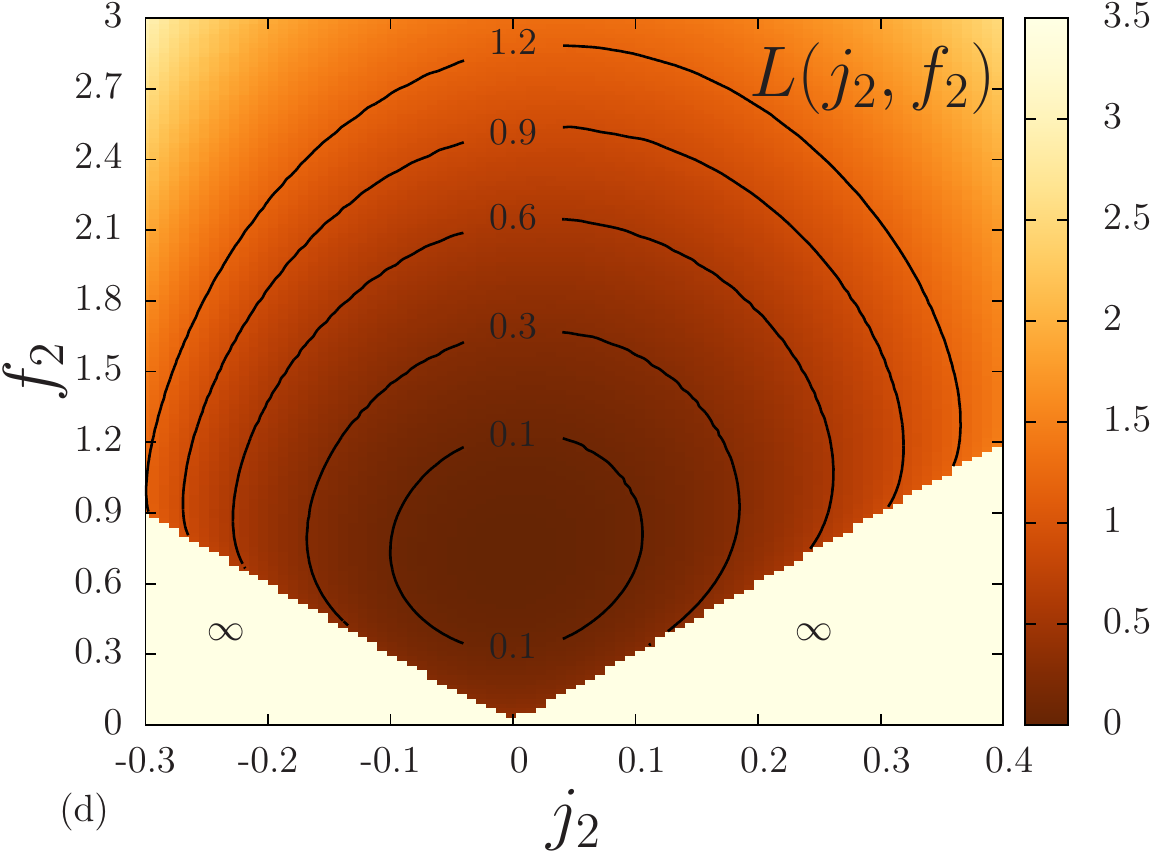} 
& \includegraphics[width=6cm]{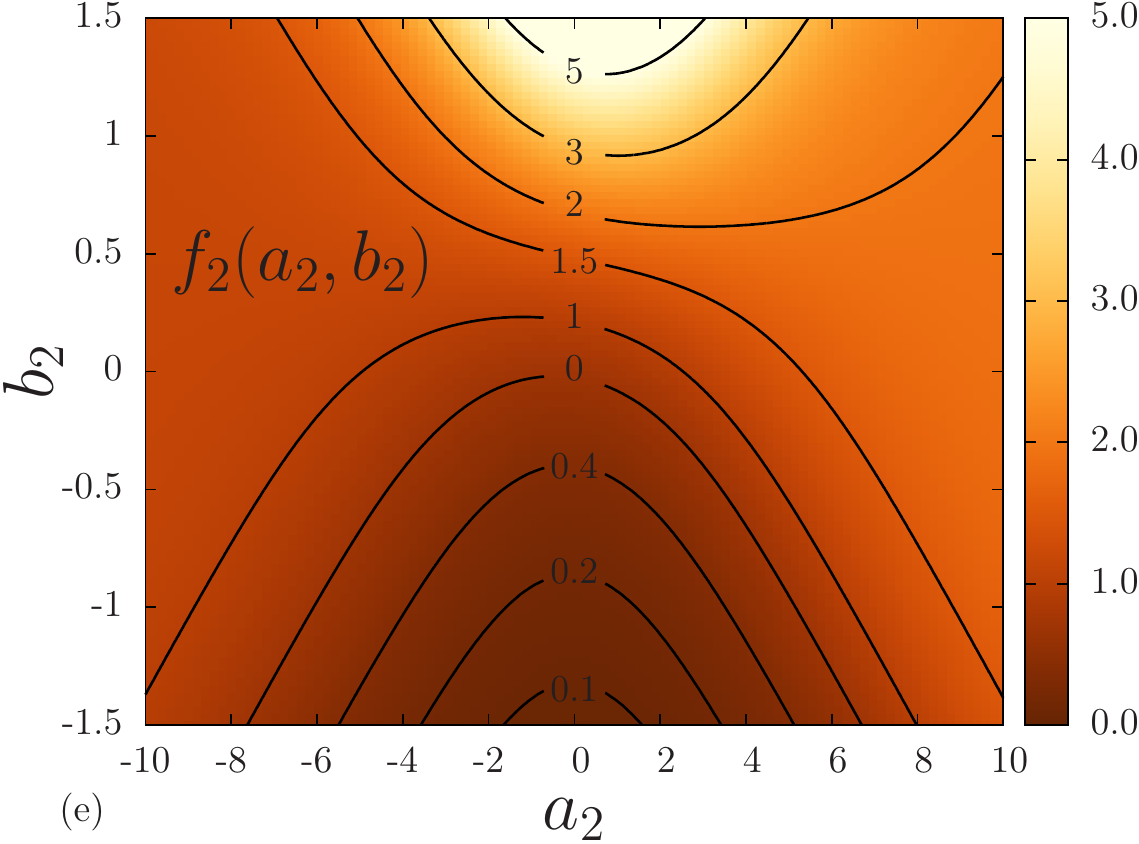} & \includegraphics[width=6cm]{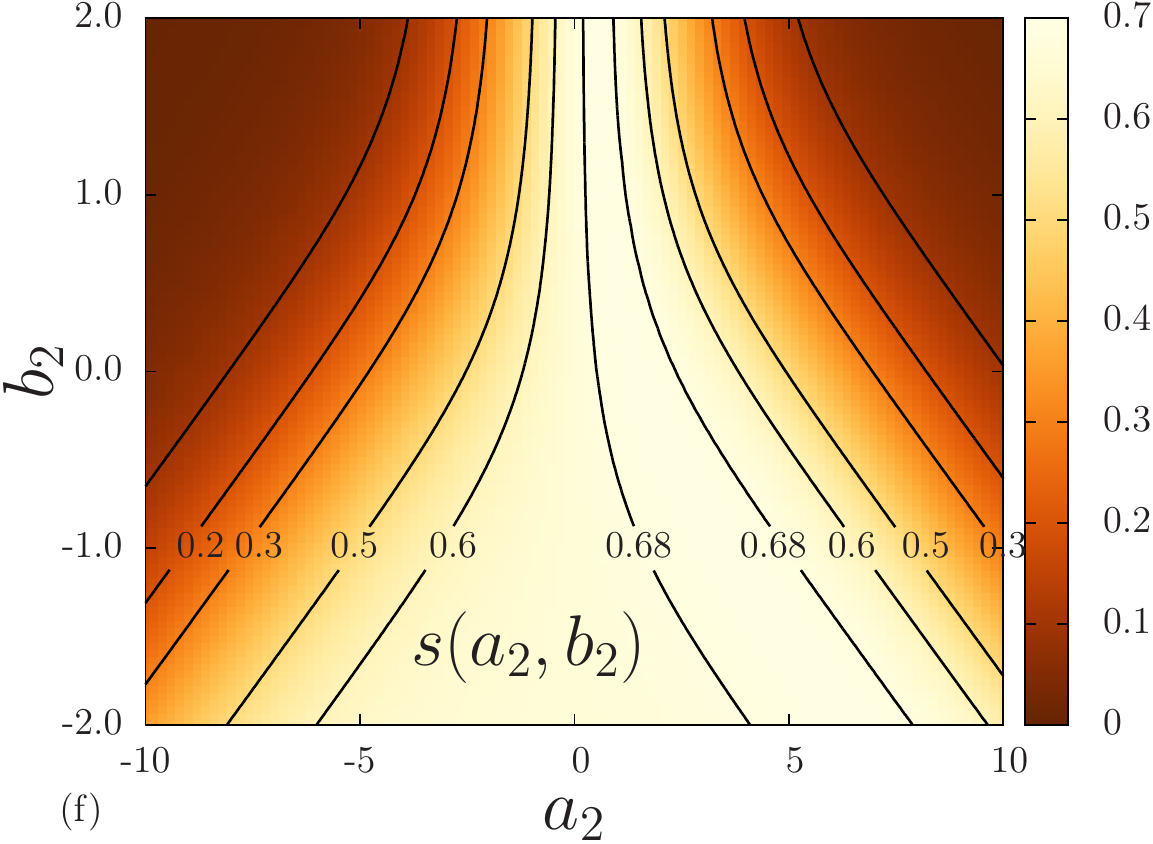}\\
\end{tabular}
\end{center}
\caption{ (a) meta-canonical potential for various $(a_2,\m_2)$ with $b=0$, (b) energy current, (c) energy, (e) activity and (f) entropy as a function of the affinity $a_2$ and the dynamical bias $b_2$. (d) NE microcanonical potentials for the energy current $j_2$ and the activity $f_2$ after a contraction on $f_1$, $p_1$ and $p_2$. Other parameters are $b_1 = 0$, $\beta_2 = \beta_1- 2 a_2 $ , $\gamma_2 = 0.5$, $\e_1=1$ and $\e_2=0.5$. For all figures, $\beta_1 = 1$ set the energy scale and $\gamma_1 =1$ the time scale. The variables $a$ and $b$ are in unit of $1/\beta_1$, the variables $j$ and $f$ are in unit of $\gamma_1/\beta_1$, and finally $L$, $\Gamma$, and $m$ are in unit of $\gamma_1$. \label{fig3}}
\end{figure*}
%%%%%%%%%%%%%
We can now illustrate the consistency of the theory: from Eqs.~(\ref{pi1}-\ref{rightEigVec}) and the product $\bpi \cdot r$, we recover the NESS probability of Eq.~(\ref{NESSprob}) obtained from derivation of the meta-canonical potential;  Eqs.~(\ref{pi1}-\ref{pi2}) allow us to compute the drift $\u$ to get the current and activity of Eqs.~(\ref{meanj}-\ref{meanf}) from Eqs.~(\ref{EqOfStatej}-\ref{EqOfStatef}) knowing the NESS probability.

We turn to the discussion of the properties of the two-level system with $\chi =2$ heat reservoirs in the light of Fig.~\ref{fig3} obtained from our analytic results. For simplicity, we chose $b_1=0$.
We set the energy scale and the time scale taking respectively $\beta_1=1$ and  $\gamma_1 =1$. Fig.~(\ref{fig3}a) shows that the meta-canonical potential is a symmetric function of the affinity $a_2$ and is strictly convex. From this symmetry, one should not conclude that the energy current $j_2$ is an anti-symmetric function of $a_2$. Indeed, the energy current comes from the derivative of the meta-canonical potential with respect to $a_2$ evaluated in $\m = c$ that has no particular symmetry when changing the sign of $a_2$.

The absolute value of the energy current $|j_2|$ and the activity $f_2$ always increases with the absolute value of the affinity $|a_2|$ at given dynamical bias $b_2$, see Fig.~(\ref{fig3}b) and  Fig.~(\ref{fig3}e). A decrease of $|j_2|$ with increasing $|a_2|$ would mean that the system has negative response for some affinities. Such a behavior is not expected for a simple two-level model. Another general trend is that $|j_2|$ and $f_2$ increases with $b_2$. Indeed, a higher dynamical bias increases the value of the transition rates corresponding to $\nu=2$, if one has $\e_+>0$, see Eq.~(\ref{ExNEProcess}). Then, a high dynamical bias accelerates the dynamics associated to reservoir $\nu=2$, whereas a small one slows it down, letting the reference dynamics associated to reservoir $\nu=1$ dominates in the transition rate matrix. Therefore, in the limit of low dynamical bias with respect to the affinity, the system approaches the EQ state at temperature $\beta_1$, with current $j_2$ and activity $f_2$ decreasing to zero.

To represent the NE micro-canonical potential $L$, one has to focus on the statistics of some specific variables by contraction: this step consists in evaluating the NE micro-canonical potential at the mean value of the disregarded variables, for instance $f_1$, $p_1$ and $p_2$ in the case of Fig.~(\ref{fig3}d). However, it is much more convenient to obtain $L(j_2,f_2)$ directly from a parametric plot of $(j_2,f_2,L)$ with $(a_2,b_2)$ being the parameters and taking $b_1=0$. In this way, we have obtained Fig.~(\ref{fig3}d) showing the NE micro-canonical potential as a convex function of $(j_2,f_2)$. This function is undefined in the regions corresponding to low activities in comparison to the energy current. The explanation is that a current can only flow if some minimal activity holds, i.e. if the system changes state regularly enough. 

Finally, the system energy $e$ and Shannon entropy $s$ are, in our framework, functions of the affinity and dynamical bias. We see in Fig.~(\ref{fig3}c) and (\ref{fig3}f) that these functions have a very similar shape in a large area corresponding to the EQ limit. The dimensionless free energy of the reference system at temperature $\beta_1$ is $\varphi = \beta_1 e-s $ and should reach its minimum value for low affinity $|a_2|$ or low dynamical bias $b_2$. There, since $\beta_1=1$, the system energy and entropy differ only in the value of the dimensionless free energy of the EQ reference system. On the contrary, at high affinity $|a_2|$, most of the time the system is either in energy state $\e_1=1$ for positive $a_2$, or $\e_2=0.5$ for negative $a_2$. The system is driven to a state where the entropy is lower than at EQ and the NE mean energy is moved away from the EQ mean value for the reference process.

\section{Conclusion}

In this paper, we have established that the asymptotic probability of the energy currents, the activities and the occupations in a NE process proceeds from the long-time statistics of the same variables at EQ. We have connected the affinities of the NE process, the dynamical biases and the escape-rate changes to constraints imposed on the EQ reference process, respectively on the energy currents, on the activities and on the occupations of each state. This connection is the analog of the ensemble equivalence between the canonical and micro-canonical ensembles of EQ statistical physics for which the temperature of the heat reservoir is associated to an energy constraint. We have argued that the mapping between EQ and NE fluctuations allows us to distinguish the reduced set of variables which play a key role in the description of NESSs.

Beyond the understanding of the structure of NE statistical physics, phenomenological and/or operational methods must be developed to compute the NE potentials of real complex systems. In this regard, it was shown that efficient algorithms exist to compute the scaled cumulants of currents \cite{Wachtel2015_vol92} or to simulate samples of rare trajectories \cite{Giardina2011_vol145}. A promising technique for macroscopic systems relies on the saddle point approximation of a path integral producing the cumulant generating function \cite{Book_Ross2008}. This calculation leads to a dynamical problem with a small number of degrees of freedom compared to the original problem. Solving this dynamical problem seems easier than finding the highest eigenvalue of a large tilted operator.

\section*{Acknowledgement} 

I thank M. Polettini and D. Lacoste for their useful comments on the manuscript and U. Dimitrijevi\'c for her careful rereading. I also thank A. Engel and A. Ran\c con for the enlightening discussions concerning respectively the large deviation theory (Appendix \ref{LDFoccupationandjump}) and the calculation of asymptotic probabilities (Appendix \ref{AppendixCountStat}).

\appendix

\section{LDF of empirical occupation and transition probability}
\label{LDFoccupationandjump}

We derive in this appendix the LDF of transition probability and occupation $I(\bo,p)$ for the EQ process. By definition, the probability that $\bo[z] = \bo$ and $p[z] =p$ when the system trajectories are generated by the EQ reference process is
\bea
P_t(\bo,p) &=& \sum_{[z]} \P_{\bk}[z] \delta_{p,p[z]} \delta_{\bo,\bo[z]}, \\
&=& \sum_{[z]} \P_{\bo/p}[z] e^{-\mathcal{A}[z]} \delta_{p,p[z]} \delta_{\bo,\bo[z]}, \label{AsympProba1}
\eea
where $\sum_{[z]}$ is the sum over all path $[z]$. We have introduced the action
\begin{equation}
\mathcal{A}[z] =\ln \frac{\P_{\bo/p}[z]}{\P_{\bk}[z]},
\end{equation}
and the path probabilities with given initial state $z(0)$
\bea
\P_{\bk}[z] &=&  \exp \left(- \int_0^t \D \tau \!\!\! \sum_{x\neq z(\tau),\nu} \!\!\! k^{\nu}_{xz(\tau)} \right) \!\!\! \prod_{\tau\in [0,t]} \!\!\! k^{\nu(\tau)}_{z(\tau+d\tau)z(\tau)}, \nonumber \\
\P_{\bo/p}[z] &=&  \exp \left(  -\int_0^t \D \tau \!\!\! \sum_{x\neq z(\tau),\nu} \!\!\! \frac{\omega^{\nu}_{xz(\tau)}}{p_{z(\tau)}} \right) \!\!\! \prod_{\tau\in [0,t]} \!\!\! \frac{\omega^{\nu(\tau)}_{z(\tau+d\tau)z(\tau)}}{p_{z(\tau)}}. \nonumber
\eea
Notice that the second line is identical to the first line where the empirical transition rate matrices $\bo^\nu[z]/p[z]$ replace the real EQ rate matrices $\bk^\nu$. From these path probabilities, the action becomes
\bea
\mathcal{A}[z] &=& \int_0^t \D \tau \sum_{x\neq z(\tau),\nu} \left( k^{\nu}_{xz(\tau)}-\frac{\omega^{\nu}_{xz(\tau)}}{p_{z(\tau)}} \right) \nonumber \\
&&+\sum_{\tau\in [0,t]} \ln \frac{\omega^{\nu(\tau)}_{z(\tau+d\tau)z(\tau)}}{k^{\nu(\tau)}_{z(\tau+d\tau)z(\tau)} p_{z(\tau)}} ,
\eea
or equivalently when introducing the empirical transition probabilities and occupations 
\bea
\mathcal{A}[z] &=& \int_0^t \D \tau \sum_{y,x\neq y,\nu} \delta_{y,z(\tau)} \left( k^{\nu}_{xy}-\frac{\omega^{\nu}_{xy}}{p_y} \right) \nonumber \\
&&+\sum_{\tau\in [0,t]} \sum_{y,x\neq y,\nu} \delta_{x,z(\tau+d\tau)} \delta_{y,z(\tau)}\delta_{\nu,\nu(\tau)} \ln \frac{\omega^{\nu}_{xy}}{k^{\nu}_{xy} p_{y}}, \nonumber\\
&=& t \sum_{x,y\neq x,\nu} \left [ p_y[z] \left( k^{\nu}_{xy}-\frac{\omega^{\nu}_{xy}}{p_y} \right) + \omega_{xy}[z] \ln \frac{\omega^{\nu}_{xy}}{k^{\nu}_{xy} p_{y}} \right]. \nonumber \\
\eea
Then, using the Kronecker symbols in Eq.~(\ref{AsympProba1}) one can take $p[z]=p$ and $\bo[z]=\bo$ to move the action out of the sum and write
\begin{equation}
P_t(\bo,p) = e^{-t I(\bo,p)} \sum_{[z]} \P_{\bo/p}[z]  \delta_{p,p[z]} \delta_{\bo,\bo[z]}. \label{AsympProba2}
\end{equation}
The remaining sum over all paths $[z]$ is the probability that $p[z]$ and $\bo[z]$ take their typical value since the path probability is generated by $\bo/p$. We expect this probability to converge to $1$ in the long-time limit. Then, we get the asymptotic probability of transitions and occupations of Eqs.~(\ref{AsymPJumpOccup}-\ref{LDF:OccupJump}) in the main text.

\section{Pre-exponential factor for the asymptotic probability of energy currents, activities and occupations}
\label{AppendixCountStat}

Following, we provide an asymptotic approximation of the long-time probability of the energy currents, activities and occupations when the final state is known. The variables are defined by
\bea
j_\nu[z]  &=& \frac{1}{t}\sum_{\tau \in [0,t]} [\e_{z(\tau+d\tau)}-\e_{z(\tau)}] \delta_{\nu ,\nu(\tau)}, \\
f_\nu[z] &=&  \frac{1}{t}\sum_{\tau \in [0,t]} d_{z(\tau+d\tau) z(\tau)} \delta_{\nu ,\nu(\tau)}, \\
p_x[z] &=& \frac{1}{t} \int_0^t \D \tau \delta_{x,z(\tau)}.
\eea 
The corresponding generating function with given final state $x$ is by definition
\begin{equation}
g_x(a,b,\m)\equiv\l \delta_{xz(t)} e^{ t\left(a^\dag \cdot j[z] + b^\dag \cdot f[z] + \m^\dag \cdot p[z]\right)} \r_{\bk} .
\end{equation}
It satisfies the linear equation $ \partial g / \partial t =\bkappa \cdot g $ with $ \bkappa $ the tilted operator defined in the main text in Eqs.~(\ref{DressedOperatorDiag}-\ref{DressedOperator}). We now look for a long-time asymptotic approximation of $ g $
\bea
g_x(a,b,\m) &=&  \left(e^{\bkappa(a,b,\m) t} \cdot p^0 \right)_{x},   \\
&\underset{t \rightarrow \infty}{\simeq} & \sum_{y}  e^{\Gamma(a,b,\m) t} \left( r  \cdot \pi^\dag  \right)_{xy} p^0_y,  \\
&\underset{t \rightarrow \infty}{\simeq}& e^{\Gamma(a,b,\m) t} r_{x}(a,b,\m) (\pi^\dag \cdot p^0), \label{AssympGenFunc}
\eea
where $p^0$ is the initial state probability. We remind that $\pi$ and $r$ are respectively the left and right eigenvectors of $\bkappa$ for the highest eigenvalue $\Gamma$. Using an asymptotic approximation to compute the inverse Laplace transformation of $ g_x(a,b,\m) $, one recovers the exponent appearing in Eq.~(\ref{ProbaAsymptotic}). 

Then, the pre-exponential factor in Eq.~(\ref{AssympGenFunc}) must be evaluated in $a=a(j,f,p)$, $b=b(j,f,p)$ and $\m=\m(j,f,p)$. We now assume that $j,f$ and $p$ are related to each other via the physical system subspace constraint, see table \ref{summary}. From Eq.~(\ref{PimappingNEtoEQ}), we find $ \pi_x(a,b,c) = \bar \pi_x(0,0,0) = 1 $ for all $x$, where the second equality stands from the fact that the left eigenvector of a Markov matrix has all its components equal to one. Then, $\pi^\dag \cdot p^0 =1$ by normalization of $p^0$ and the right eigenvector of $\kappa$ in the physical system subspace is the NESS probability $r(a,b,c) = p^*$ for the dynamics with energy current $j$ and activity $f$. This leads to the asymptotic probability of energy currents, activities and occupations when the final state at time $t$ is $x$
\begin{equation}
P_t(j,f,p^*,x) \underset{t \ra \infty}{\simeq} e^{ -t L(j,f,p^*)}p^*_x. 
\end{equation}

\section{Asymptotic equivalence of the NE process and the canonical process}

\label{EnsembleEquivalence2}

The path probability of the canonical process, with generator $\bcalK$, is defined by exponentially weighting the path probability of the EQ reference process:
\beq
\P_{\bcalK}[y] \equiv \frac{\P_{\bm{k}}[y] e^{ t\left(a^\dag \cdot j[y] + b^\dag \cdot f[y] + \m^\dag \cdot p[y]\right)} }{\l e^{ t\left(a^\dag \cdot j[z] + b^\dag \cdot f[z] + \m^\dag \cdot p[z]\right)} \r_{\bk}}
\eeq
This tilting procedure is sometime referred to as canonical conditioning. We show in this section that the above canonical process is asymptotically equivalent to the NE process defined in Sec.~\ref{DefNEprocess}. From Sec.\ref{EnsembleEquivalence}, it is also equivalent to the driven process. The connection between the driven process and the canonical process was first obtained in Refs.~\cite{Jack2010_vol184, Garrahan2010_vol104} and studied in depth in Refs.~\cite{Chetrite2015_vol16, Chetrite2015_vol2015}.

From the definition of the CGF in Eq.~(\ref{DefCGF}), we have
\beq
\P_{\bcalK}[y] \underset{t \rightarrow \infty}{\simeq} \P_{\bm{k}}[y] e^{ t\left(a^\dag \cdot j[y] + b^\dag \cdot f[y] + \m^\dag \cdot p[y]\right) - t \Gamma }. \label{AsympPcano}
\eeq
Since Eq.~(\ref{exitrule}) is satisfied for all $y$, we can write it for any state $y(\tau)$ along the trajectory $[y]$ 
\beq
\Gamma = \m_{y(\tau)} +  \Lambda_{y(\tau)} -  \lambda_{y(\tau)},
\eeq
and upon integration over the time $\tau$, one finds
\beq
t \Gamma = t m^\dag \cdot p[y] + \int_0^t \D\tau (\bar \lambda_{y(\tau)} -\lambda_{y(\tau)} ), \label{tgamma}
\eeq
since $\Lambda = \bar \lambda $ from Eq.~(\ref{DrivEQandNEescapeEquality}). Finally, Eqs.~(\ref{AsympPcano}-\ref{tgamma}) leads to the asymptotic equivalence of the path probabilities 
\bea
\P_{\bcalK}[y] & \underset{t \rightarrow \infty}{\simeq}& e^{- \int_0^t \D\tau \bar \lambda_{y(\tau)} + t\left(a^\dag \cdot j[y] + b^\dag \cdot f[y] \right) } \!\!\! \prod_{\tau\in [0,t]} \!\!\! k^{\nu(\tau)}_{y(\tau+d\tau)y(\tau)}, \nonumber \\
& \underset{t \rightarrow \infty}{\simeq}& e^{- \int_0^t \D\tau \bar \lambda_{y(\tau)}  } \!\!\! \prod_{\tau\in [0,t]} \!\!\! \bar k^{\nu(\tau)}_{y(\tau+d\tau)y(\tau)} \\
& \underset{t \rightarrow \infty}{\simeq}& \P_{\bar \bk}[y].
\eea
Hence, we have proved that the canonical process corresponds at long time to a NE process that can be realized experimentally changing the temperatures of the heat reservoirs and the dynamical biases.

\section{Fluctuation Theorem \label{FT}}

The FT is an essential property of the stochastic entropy production \cite{Seifert2012_vol75,VandenBroeck2014_vol418}. According to this theorem, a stochastic positive entropy production is exponentially more likely than the opposite entropy production, i.e. an entropy destruction. On average this implies a positive entropy production in agreement with the second law. Therefore, the FT is a probabilistic statement of the second law, and as such it is a very fundamental property of NE phenomena. It was first derived with a long-time approximation, but since the mean entropy production always increases, a FT should hold at all time \cite{BulnesCuetara2014_vol,Polettini2014_vol2014}. Because the entropy production may be appropriately defined using different NE variables such as work, heat or particle currents depending on the experimental setup, the FT has many faces \cite{Jarzynski1997_vol78, Kurchan1998_vol31, Crooks2000_vol61, Andrieux2004_vol121, Seifert2005_vol95, Sagawa2010_vol104, Esposito2010_vol104, Verley2012_vol108, Crisanti2013_vol110}. Generally, the joint probability distribution of a set of time anti-symmetric variables summing to entropy production will satisfy a FT \cite{Garcia-Garcia2010_vol82}. In our case, a linear combination of the currents gives the entropy production rate
\begin{equation}
\s =  2 a^\dag \cdot j .
\end{equation}
Accordingly, the LDF and scaled CGF for the NE process have a FT symmetry. This symmetry strongly relies on local detailed balance, in other word, on the symmetry of transition rates. We already used the local detailed balance to show the equivalence of EQ and NE fluctuations. We show in this appendix that the fluctuation theorem (FT) is a consequence of the mapping between EQ and NE fluctuations associated to the symmetric nature of energy-currents fluctuations at EQ. Using Eq.~(\ref{CGFmappingNEtoEQ}), we find
\bea
\bar \Gamma(-2a-\bar a,\bar b,\bar \m) &=&  \Gamma(-2a-\bar a+a,\bar b+b,\bar \m+c)  \nonumber \\
&=& \Gamma(\bar a+a,\bar b+b,\bar \m+c\bar b+b,\bar \m+c) \nonumber \\
&=& \bar \Gamma(\bar a,\bar b,\bar \m),
\eea
where we have used the fact that EQ fluctuations are symmetric in the reversal of affinities $ \Gamma(a,b,\m) = \Gamma(-a,b,\m) $. Similarly, from Eq.~(\ref{LDFmappingNEtoEQ}), it is straightforward to see that
\begin{equation}
\bar L(j,f,p)- \bar L(-j,f,p)= - 2a^\dag \cdot j = - \s,
\end{equation}
since we have $ L(j,f,p)= L(-j,f,p)$.

\bibliography{Ma_base_de_papier}

\begin{thebibliography}{80}
\bibitem [{1()}]{1}%
  \BibitemOpen
  \href@noop {} {}\bibinfo {note} {Beyond the case of energy sources, one can
  imagine an experimental setup where a matter flux or a rate of volume
  increase can be controlled exactly.}\BibitemShut {Stop}%
\bibitem [{tim()}]{time}%
  \BibitemOpen
  \href@noop {} {}\bibinfo {note} {We have multiplied the covariances by a
  factor $t$ since we consider time-averaged variables; we would have divided
  by $t$ if we were considering the time-integrated variables.}\BibitemShut
  {Stop}%
\end{thebibliography}

\end{document}